\documentclass[12pt]{iopart}
\usepackage{graphicx}
\begin{document}

\title{Random-field-induced order in bosonic t-J model}

\author{Yoshihito Kuno, Takamasa Mori, and Ikuo Ichinose}

\address{Department of Applied Physics, Nagoya Institute of Technology, 
Nagoya, 466-8555, Japan}
\begin{abstract}
In the present paper, we shall study effect of a random quenched external
field for spin order and also multiple Bose-Einstein condensation (BEC).
This system is realized by the cold atomic gases in an optical lattice.
In particular, we are interested in the strong-repulsion region of two-component
gases for which the bosonic t-J model is a good effective model.
In the bosonic t-J model, a long-range order of the pseudo-spin and also BEC
of atoms appear quite naturally as in the fermion t-J model for the 
high-temperature superconducting materials.
Random Raman scattering between two internal states of a single atom 
plays a role of the random external field, and we study its effects on 
the pseudo-spin order and the BEC by means of quantum Monte-Carlo simulations. 
The random external field breaks a continuous U(1)
symmetry existing in the original bosonic t-J model and it induces new orders
named random-field-induced order (RFIO).
We show a phase diagram of the bosonic t-J model with the random 
external magnetic field and study the robustness of the RFIO states.
We also study topological excitations like vortices and domain wall in the
RFIO state.
Finally we point out the possibility of a quantum bit by the RFIO.
\end{abstract}

%\maketitle

%%%%%%%%%%%%%%%%%%%%%%%%%%%%%%%%%%%%%%%%%%%%%%%%%%%%%%
\section{Introduction}

Quenched disorder plays very important role in condensed matter physics.
A prominent example is the Anderson localization that predicts all quantum states are localized in one and two spatial dimensions if interactions between particles
can be neglected\cite{anderson}.
It has been proved rigorously that quenched disorder destroys
ordered states and rounds singularity of phase transitions\cite{azenman}.
However recently, a counter-intuitive possibility was pointed out and
examined.
That is, a quenched disorder generates a new ordered state,
which is different from the original one,
if that quenched disorder breaks a continuous symmetry of the original system
without the quenched disorder.
This possibility was first studied in a classical XY spin model coupled with a
random external magnetic field\cite{RFIO1} and then
the resultant order is called random-field-induced order (RFIO).
Shortly after the proposal, it was shown that such phenomena of the
RFIO can be observed by experiments on systems of ultra-cold bosonic atoms 
of multiple-internal states\cite{RFIO2}.
In Bose atomic gases, a Bose-Einstein condensed (BEC) state is the 
genuine ordered state,
and a random Raman scattering of the internal states of the atom plays a role of 
a quenched random external magnetic field.
Then, it is expected that the RFIO can be observed in the ultra-cold
atomic gas systems.
There are other interesting works on the RFIO\cite{RFIO3}.
 
In this paper, we shall investigate the RFIO in detail for the two-component
cold bosonic gases in a square optical lattice (OL).
In particular, we consider the bosonic t-J model, which is a low-energy
effective model for the Bose-Hubbard model in the strong-repulsion 
limit\cite{BtJ2,BtJ}.
This model exhibits both the pseudo-spin order and the BECs and is therefore
suitable for study of the RFIO.
We employ quantum Monte-Carlo (MC) simulations that take into account 
all of fluctuations. 
Thus the present study is in sharp contrast to the previous ones that used 
estimation of classical energy of the XY spin configurations
for observing a possible RFIO\cite{RFIO1} and 
a mean-field theory with Gross-Pitaevskii equations for the BEC
of the RFIO\cite{RFIO2}. 
Furthermore we will investigate behavior of low-energy topological excitations
such as vortices and domain walls in the RFIO states, and reveal interesting
properties of them.
Finite-temperature ($T$) phase diagram is also obtained, which is useful for discussion
on the robustness of the RFIO states.

The present paper is organized as follows.
In Sec.2, we introduce and explain the bosonic t-J model with a random external field.
Path-integral quantization using the slave-particle representation is
explained.
Effective field theory for the pseudo-spin and BECs is derived by
integrating out amplitude degrees of freedom of the slave-particle
field variables.
In Sec.3, a replica mean-field theory is applied to the effective field theory
and effects of the random field are studied.
This study clearly shows how the RFIO appears as a result of the
random field with a moderate fluctuation. 
In Sec.4, the results of the numerical simulations are given.
Phase diagrams at vanishing $T$ as well as finite-$T$ are obtained.
Various correlation functions, which are used for the identification of the orders
are shown.
In Sec.5., topological excitations like vortices and domain wall are
investigated numerically.
Properties of these excitations are discussed from the view point of the
RFIO.
Section 6 is devoted for conclusion.

%%%%%%%%%%%%%%%%%%%%%%%%%%%%%%%%%%%%%%%%%%%%%%%%%%%%%%
\section{Models and numerical methods}

\subsection{Bosonic t-J model with random Rabi coupling} 

The bosonic t-J model, which describes dynamics of two internal states of a boson,
which we call $a$- and $b$-boson for simplicity,
in a square OL, is defined by the following Hamiltonian\cite{BtJ},
\begin{eqnarray}
&&H_{\rm EtJ}=H_{\rm tJ}+H_V, \label{HEtJ} \\
&&H_{\rm tJ}=-\sum_{\langle i,j\rangle} (t_a a^\dagger_{i}a_j
+t_b b^\dagger_{i}b_j+\mbox{h.c.})  
-J_{xy}\sum_{\langle i,j\rangle}(S^x_{i}S^x_j+S^y_{i}S^y_j)  \nonumber \\
&&+J_z\sum_{\langle i,j\rangle}S^z_iS^z_j
-\mu\sum_i(1-a^\dagger_ia_i-b^\dagger_ib_i), \label{HtJ}  \\ 
&&H_V={V_0 \over 4}\sum_i\Big((a_i^\dagger a_i-\bar{\rho}_{ai})^2+
(b_i^\dagger b_i-\bar{\rho}_{bi})^2\Big),
\label{HV}
\end{eqnarray}
where $a^\dagger_i(a_i)$ and $b^\dagger_i(b_i)$ are 
boson creation (destruction) operators at site $i$ of the square lattice 
and $t_a$ and $t_b$ are the hopping amplitude between the nearest-neighbor (NN)
sites.
Pseudo-spin operator $\vec{S}_i$ is given as  
$\vec{S}_i={1 \over 2}B^\dagger_i\vec{\sigma}B_i$ with
$B_i=(a_i,b_i)^t$, and $\vec{\sigma}$ is the Pauli spin matrix.
In the t-J model, the doubly-occupied state is excluded at each site
and density of atoms is controlled by the chemical potential $\mu$.
It was probed that the Hamiltonian $H_{\rm tJ}$ is derived from the Bose-Hubbard 
model in the strong one-site repulsion limit by integrating out multiple-particle states,
and the exchange couplings 
$J_{xy}$ and $J_z$ are related with the intra and inter-repulsions between
atoms.
In the present study, however, we shall treat these parameters as free ones
because the system $H_{\rm tJ}$ might be derived from a Bose-gas system on the
Lieb lattice that is a bosonic counterpart of the d-p model for the 
strongly-correlated electron systems. 

The term $H_V$ in Eq.(\ref{HV}) controls density fluctuations of atoms at each site
from the mean value $\bar{\rho}_{ai}$ and $\bar{\rho}_{bi}$.
This term is expected to appear naturally for describing practical phenomena
in experiments at low energies and therefore we explicitly added it to the
Hamiltonian.

In the later discussion, we shall mostly consider the case $J_z=0$, which corresponds to
the case of the equal intra and inter-species repulsions in the Bose-Hubbard model.
The bosonic t-J model without the random external field was studied 
by both the numerical and analytical methods in the previous 
papers and its phase diagram has been clarified\cite{BtJ,BtJ2}.
For sufficiently large $J_{xy}$, a ferromagnetic (FM) state of the pseudo-spin 
appears, whereas for sufficiently large hopping amplitude $t_a$ and $t_b$,
BECs of the atoms form.
As the anti-ferromagnetic coupling $J_z$ is increased, supersolid forms
for a small but finite parameter region.
For sufficiently large $J_z$ compared with $t_a, \; t_b$ and $J_{xy}$, a solid state
with the checkerboard density pattern appears as the lowest-energy state.

In the present study, we add the following terms that describe the quenched random
external fields,
\begin{equation}
H_{\cal T}=H_{\rm EtJ}+\sum_i(J^x_iS^x_i+J^y_iS^y_i),
\label{HT}
\end{equation}
where $J^x_i$ and $J^y_i$ take random real variables with the vanishing mean value.
For the practical numerical study, we use the following distribution function $P(J_i)$,
\begin{equation}
P(J^x_i)={1 \over \sigma_x\sqrt{\pi}}\ \exp [-(J^x_i/\sigma_x)^2], \;\;
P(J^y_i)={1 \over \sigma_y\sqrt{\pi}}\ \exp [-(J^y_i/\sigma_y)^2],
\label{PJ}
\end{equation}
where $\sigma_{x(y)}$ are positive parameters.
In the cold atomic systems, the above terms are realized by the  
Rabi oscillation with the Raman laser of a random complex amplitude 
$\Omega_i=\Omega^{\rm R}_i+i\Omega^{\rm I}_i$, and then
$J^x_i(J^y_i) \propto \Omega^{\rm R}_i(\Omega^{\rm I}_i)$ as 
$S^x_i=a^\dagger_ib_i+b^\dagger_ia_i$ and 
$S^y_i={1 \over i}(a^\dagger_ib_i-b^\dagger_ia_i)$.
It is expected that the complex Raman amplitude $\Omega$ is realized experimentally
using speckle laser light\cite{raman}. 

It should be remarked that the U(1)$\times$U(1) symmetry of 
$H_{\rm EtJ}$ in Eq.(\ref{HEtJ}), i.e.,
$(a_i,b_i)\rightarrow (e^{i\alpha}a_i,e^{i\beta}b_i)$ with arbitrary 
constants $\alpha$ and $\beta$,
is preserved only for the case $\sigma_x=\sigma_y$, otherwise the quenched
disorder, $J^x_i$ and $J^y_i$, explicitly breaks the symmetry as 
U(1)$\times$U(1)$\rightarrow $U(1)$\times Z_2$.

%%%%%%%%%%%%%%%%%%%%%%%%%%%%%%%%%%%%%%
\subsection{Numerical methods:Path-integral Monte-Carlo simulations} 

In order to study the model $H_{\cal T}$ by means of quantum MC simulations,
we use the path-integral method with the slave particle description for
the local constraint of the t-J model.
The boson creation operators are expressed by the slave particle
operators $\phi_{ai}, \phi_{bi}$ and $\phi_{hi}$ as follows,
\begin{equation}
a^\dagger_i=\phi^\dagger_{ai}\phi_{hi},  \;\;\;
b^\dagger_i=\phi^\dagger_{bi}\phi_{hi},
\label{slave}
\end{equation}
and physical state of the slave particle $|Phys\rangle$ must satisfy
\begin{equation}
(\phi^\dagger_{ai}\phi_{ai}+\phi^\dagger_{bi}\phi_{bi}+\phi^\dagger_{hi}\phi_{hi})
|Phys\rangle=|Phys\rangle.
\label{const}
\end{equation}
Then the partition function for the system $H_{\cal T}$ is given by
\begin{equation}
Z=\int[d\phi_ad\phi_bd\phi_h]\exp\Big[-\int d\tau \Big(
\sum_{\alpha=a,b,h}(\bar{\phi}_{\alpha i}\dot{\phi}_{\alpha i})
+H_{\cal T}\Big)\Big],
\label{Z}
\end{equation}
where $\tau$ is the imaginary time, $\dot{\phi}_{\alpha i}={d\phi_{\alpha i} \over d\tau}$
and $H_{\cal T}$ is expressed in terms of the slave particles by using Eq.(\ref{slave}).
For the path integral in Eq.(\ref{Z}), the local constraint 
$\bar{\phi}_{ai}\phi_{ai}+\bar{\phi}_{bi}\phi_{bi}+\bar{\phi}_{hi}\phi_{hi}=1$
can be imposed by using a Lagrange multiplier field $\lambda_i(\tau)$,
\begin{eqnarray}
\prod_\tau
\delta(\bar{\phi}_{ai}\phi_{ai}+\bar{\phi}_{bi}\phi_{bi}+\bar{\phi}_{hi}\phi_{hi}-1) =\int[d\lambda]e^{i\int d\tau
(\bar{\phi}_{ai}\phi_{ai}+\bar{\phi}_{bi}\phi_{bi}+\bar{\phi}_{hi}\phi_{hi}-1)\lambda_i}.
\end{eqnarray}

To obtain a positive-definite action for carrying out the path-integral MC simulation,
we parameterize the fields as 
$\phi_{\alpha i}=\sqrt{\rho_{\alpha i}} \ e^{i\omega_{\alpha i}}$ $(\alpha=a,b,h)$, and 
analytically calculate the integral over the amplitudes $\rho_{\alpha i}$.
By the term $H_V$ in $H_{\cal T}$, the integration can be carried out in
powers of the density fluctuations, 
$\delta\rho_{\alpha i}=\rho_{\alpha i}-\bar{\rho}_{\alpha i}$.
As a result, the Berry phase 
$\sum_\alpha\bar{\phi}_{\alpha i}\dot{\phi}_{\alpha i}$ generates terms like 
${1 \over V_0}\sum_\alpha (\dot{\omega}_{\alpha i}+\lambda_i)^2$ in the action.

For the practical numerical calculation, we introduce a lattice for the imaginary
time direction, and we denote a site of three-dimensional (3D) cubic 
space-time lattice $r$.
With the resultant action on the lattice $A_{\rm Lxy}$, the partition function
is given by
\begin{equation}
Z=\int [d\omega_{\alpha r} d\lambda_r]e^{-A_{\rm Lxy}},
\label{Z2}
\end{equation} 
with
\begin{equation}
A_{\rm Lxy}=A_{{\rm L}\tau}+A_{\rm L}(e^{i\theta_\sigma},e^{-i\theta_\sigma})
+A_{\rm q}, 
\label{AL1}
\end{equation}
where
\begin{eqnarray}
A_{{\rm L}\tau}=-c_\tau\sum_{r} \sum_{\alpha=a,b,h}
\cos (\omega_{\alpha,r+\hat{\tau}}-\omega_{\alpha r}+\lambda_r),  
\label{Atau}
\end{eqnarray}
%and 
\begin{eqnarray}
A_{\rm L}(e^{i\theta_\sigma},e^{-i\theta_\sigma}) &=& 
-\sum_{\langle r,r'\rangle}
\Big(C^a_3\cos (\theta_{ar}-\theta_{ar'})+
C^b_3\cos (\theta_{br}-\theta_{br'}) \nonumber \\
&&+C_1\cos (\theta_{sr}-\theta_{sr'})\Big),
\label{AL2}
\end{eqnarray}
and
\begin{eqnarray}
A_{\rm q} =
-\sum_{\langle r,r'\rangle}
\Big(\tilde{J}^x_i\cos (\theta_{sr}-\theta_{sr'})+
\tilde{J}^y_i\sin (\theta_{sr}-\theta_{sr'})\Big).
\label{Aq}
\end{eqnarray}
In Eqs.(\ref{AL1}) $\sim$ (\ref{Aq}), dynamical variables are 
$$
\theta_{s r}=\omega_{a r}-\omega_{b r}, \
\theta_{a r}=\omega_{a r}-\omega_{h r}, \
\theta_{b r}=\omega_{b r}-\omega_{h r},
$$
and parameters are related to the original ones as,
\begin{eqnarray}
&& c_\tau={1 \over V_0\Delta\tau},   \nonumber \\
&& C_1=4J\bar{\rho}_a^2\bar{\rho}_b^2\Delta \tau 
\propto {J/(c_\tau V_0)}, \nonumber  \\
&& C^a_3={t_a \over 2}\bar{\rho}_a(1-\bar{\rho}_a-\bar{\rho}_b)\Delta \tau  
\propto {t_a/(c_\tau V_0)},  \nonumber \\
&& C^b_3={t_b \over 2}\bar{\rho}_b(1-\bar{\rho}_a-\bar{\rho}_b)\Delta \tau  
\propto {t_b/(c_\tau V_0)},  \nonumber \\
&& \tilde{J}^{x(y)}={J}^{x(y)}\Delta \tau={J}^{x(y)}/(c_\tau V_0),  
\label{parameters}
\end{eqnarray}
where $\Delta\tau$ is the lattice spacing of the imaginary time.
Note that $c_\tau, \cdots, \tilde{J}^y$ are all dimensionless.
(We have put $\hbar=1$.)
Please notice that the quenched disorder variables $\tilde{J}^{x(y)}_i$ are
independent of the imaginary time $\tau$.

There are comments on the derivation of $A_{\rm Lxy}$ and advantages
of the MC simulation on it.
On performing the path integral of $\rho_{\alpha i}$, the higher-order terms of 
the fluctuations $\delta\rho_{\alpha i}$ are ignored, e.g., in the hopping term,
\begin{equation}
a^\dagger_i a_j \rightarrow \sqrt{\bar{\rho}_{ai}\bar{\rho}_{aj}}
\exp[i(-\omega_{ai}+\omega_{hi}
+\omega_{aj}-\omega_{hj})].
\label{appr}
\end{equation}
The above approximation is legitimate for $\delta \rho/\bar{\rho} \ll 1$ as
in the experiments of large $\bar{\rho}$\cite{BKT} or small $\delta \rho$.
In the previous paper\cite{QXY2}, we studied the non-random case rather in detail
and verified that results obtained by the MC simulations on $A_{\rm Lxy}$
are in good agreement with those obtained by the Gross-Pitaevskii theory.
Furthermore in the previous paper\cite{QXY1}, we studied the phase diagram
of $H_{\rm EtJ}$ with a finite $J_z$ by using $A_{\rm Lxy}$ and
found that the supersolid state forms in certain parameter region.
The obtained phase diagram is in agreement with that
of the two-component Bose-Hubbard model, which was obtained
by using the MC simulation with the worm algorithm\cite{2BH},
although the case of commensurate filling factors was studied there.
One advantage of the present MC method for studying the bosonic
t-J model is its rapid convergence, and therefore the large-scale
MC simulation is possible.
Furthermore, various correlation functions as well as the density
of topological excitations can be calculated accurately, as we show 
in sections 4 and  5.

%%%%%%%%%%%%%%%%%%%%%%%%%%%%%%%%%%%%%%%%%%%%%%%%%%%%%%%
\section{Replica mean-field theory}

Before going into the numerical calculations, we briefly study
the model given by Eq.(\ref{Z2}) by means of the replica methods.
In particular, we are interested in the case of the single-component
random field like $\tilde{J^x}\neq 0$ and $\tilde{J^y}=0$, and see
how the order of $S^y$ shows up whereas that of $S^x$ does not.
For simplicity, we shall consider the case of the total filling factor =1, i.e.,
the filling factor of each particle is $1/2$,
and focus on the pseudo-spin symmetry, though the extension to
the case with a finite hole density is rather straightforward.

In the replica method studying effects of quenched random variables, 
a replica index $\nu=1,2, \cdots, n$ is introduced for each dynamical variable.
In the present system, $\omega_{\alpha i}\rightarrow \omega^\nu_{\alpha i} \
(\alpha=a,b,h)$,
and the partition function of the replica system $[Z^n]$ is given by,
\begin{eqnarray}
[Z^n]&=& \int \Big\{\prod_i (d\tilde{J}^x_i)P(\tilde{J}^x_i)\Big\}
\Big\{\prod_{i,\nu} (d\omega^\nu_{ai}d\omega^\nu_{bi})\Big\}
\exp\Big[-\sum_\nu \Big(A_{\rm S}^\nu+\tilde{J}^x_i\int d\tau S^{x\nu}_{i}
\Big)\Big],  \nonumber \\
A^\nu_{\rm S}&=&\int d\tau\Big[{1 \over V_0}
\sum_{i,\nu}(\dot{\omega}^\nu_{ai}+\dot{\omega}^\nu_{bi})
-C_1\sum_{i,\mu}(S^{x\nu}_iS^{x\nu}_{i+\mu}+
S^{y\nu}_iS^{y\nu}_{i+\mu})\Big],
\label{replicaZ}
\end{eqnarray}
where $\dot{\omega}^\nu_{ai}={d\omega^\nu_{ai} \over d\tau}$, etc,
and $[\cdots]$ denotes average over the random variables $\tilde{J}^x_i$
with $P(\tilde{J}^x_i)$.
After calculating $[Z^n]$, the limit $[Z^n] \rightarrow (1+n[\log Z])$ for 
$n\rightarrow 0$ is taken to obtain $[\log Z]$.

In $[Z^n]$ in Eq.(\ref{replicaZ}), the integration over $\tilde{J}^x_i$ can be
carried out readily to obtain,
\begin{equation}
\hspace{-1.5cm}
[Z^n]= \int 
\Big\{\prod_{i,\nu} (d\omega^\nu_{ai}d\omega^\nu_{bi})\Big\}
\exp\Big[-\sum_\nu A_{\rm S}^\nu+{\sigma^2_x \over 4}
\sum_i\Big(\int d\tau \sum_\nu S^{x\nu}_{i}(\tau)\Big)
\Big(\int d\tau' \sum_{\nu'} S^{x\nu'}_{i}(\tau')\Big)
\Big].
\label{replicaZ2}
\end{equation}
The nonlocal terms in Eq.(\ref{replicaZ2}) can be reduced to local ones by 
using a Hubbard-Storatonovich transformation with auxiliary fields $m_i(\tau)$ as
\begin{eqnarray}
&&\exp\Big[{1 \over 4\sigma_x}
\sum_i\Big(\int d\tau \sum_\nu S^{x\nu}_{i}(\tau)\Big)
\Big(\int d\tau' \sum_{\nu'} S^{x\nu'}_{i}(\tau')\Big)
\Big]  \nonumber \\
&& \hspace{1.5cm} 
=\int [dm_i]\exp\Big[-{1 \over \sigma^2_x}\sum_im^2_i+\sum_im_i
\Big(\int d\tau \sum_\nu S^{x\nu}_{i}(\tau)\Big)\Big].
\label{HSt}
\end{eqnarray}
We also apply a mean-field theory (MFT) for the spin part of $A^\nu_{\rm S}$
in Eq.(\ref{replicaZ}) as
\begin{equation}
\sum_{i,\mu}(S^{x\nu}_iS^{x\nu}_{i+\mu}+
S^{y\nu}_iS^{y\nu}_{i+\mu}\Big) \rightarrow
\sum_i \Big(4\langle S^{x\nu}\rangle S^{x\nu}_i-2\langle S^{x\nu}\rangle^2
+4\langle S^{y\nu}\rangle S^{x\nu}_i-2\langle S^{y\nu}\rangle^2\Big).
\end{equation}
In this MFT, the partition function of the replica system $[Z^n]_{\rm MFT}$
is given as 
\begin{eqnarray}
[Z^n]_{\rm MFT}&=&\int [dm_i][d\omega_a][d\omega_b]
\exp\Big[-{1 \over \sigma^2_x}\sum_im^2_i+\int d\tau\sum_{i,\nu}
(m_i+4C_1\langle S^{x\nu}\rangle)S^{x\nu}_i  \nonumber \\
&&+\int d\tau \sum_{i,\nu}\{-2C_1\langle S^{x\nu}\rangle^2
+4C_1\langle S^{y\nu}\rangle S^{y\nu}_i-2C_1\langle S^{y\nu}\rangle^2\} \nonumber \\
&&-{1 \over V_0}\int d\tau \sum_{i,\nu}((\dot{\omega}^{\nu}_{a,i})^2
+(\dot{\omega}^{\nu}_{b,i})^2)\Big].
\label{Zmft}
\end{eqnarray}
In Eq.(\ref{Zmft}), the integration of $\omega^\nu_{\alpha i}$ can be 
carried out to obtain a Ginzburg-Landau theory (GL theory) for 
the pseudo-spin order.
To this end, we use the following on-site Green functions as we consider
the system at sufficiently low temperature,
\begin{equation}
\langle e^{i\omega^\nu_{\alpha, i}(\tau)} e^{-\omega^{\nu'}_{\alpha, j}(\tau')} \rangle
=\delta_{\nu\nu'}\delta_{ij}e^{-V_0|\tau-\tau'|}, \;\;\; (\alpha=a,b).
\label{Gfunc}
\end{equation}
Then
\begin{eqnarray}
[Z^n]_{\rm MFT}&=&\int [dm_i]
\exp\Big[-{1 \over \sigma^2_x}\sum_im^2_i-2C_1\int d\tau\sum_\nu
(\langle S^{x\nu}\rangle^2+\langle S^{y\nu}\rangle^2)  \\
&&+\int d\tau \sum_{i,\nu}\Big\{ {(\gamma_i^\nu)^2 \over 4V_0}
+{4C_1^2 \over V_0}\langle S^{y\nu}\rangle^2\Big\}\Big],
\end{eqnarray}
where $\gamma^\nu_i=m_i+4C_1\langle S^{x\nu}\rangle$.
We are interested in the replica-symmetric solution and set
$\sum_\nu \langle S^{x\nu} \rangle=n \langle S^{x} \rangle$, etc.
We also introduce a cutoff $\beta=1/(k_{\rm B}T)$ for the integral of 
the imaginary time $\tau$.
Then the integration over $m_i$ can be done to obtain
\begin{eqnarray}
&&\int dm_i \exp\Big[-{1 \over \sigma^2_x} m^2_i+{\beta n\over 4V_0}
(m_i+4C_1\langle S^{x}\rangle)^2\Big]  \nonumber \\
&&\hspace{1cm}
=\exp\Big[{1 \over {1 \over \sigma^2_x}-{\beta n \over 4V_0}}\Big({\beta n \over V_0}
C_1\langle S^x\rangle\Big)^2+{4C_1^2\beta n \over V_0}\langle S^x\rangle^2
\Big].
\label{mint}
\end{eqnarray}
Finally we obtain the effective potential, $V_{\rm Rep}$, by taking the limit
$n\rightarrow 0$,
\begin{eqnarray}
V_{\rm Rep}&\equiv& {\rm limit}_{n\rightarrow 0}
\Big({1-[Z^n]_{\rm MFT}\over \beta n}\Big) \nonumber  \\
&=&\Big(2C_1-{4C_1^2 \over V_0}\Big)\Big(\langle S^x\rangle^2+\langle S^y\rangle^2\Big)
-{\beta n \over {1\over \sigma^2_x}-{\beta n \over 4V_0}}\Big({C_1 \over V_0}
\langle S^x\rangle\Big)^2.
\label{Vrep}
\end{eqnarray}

It is obvious that two limits, $\beta \rightarrow \infty$ and 
$n \rightarrow 0$, are {\em not} interchangeable in $V_{\rm Rep}$
given by Eq.(\ref{Vrep}).
For a finite $\sigma_x$ and at finite temperature, 
the last term in $V_{\rm Rep}$ (\ref{Vrep})
vanishes for $n\rightarrow 0$ and both $\langle S^x \rangle$ and $\langle S^y \rangle$
can have a nonvanishing value for $C_1 >{V_0 \over 2}$, i.e., in the case in which the 
spin interaction $J$ dominates the suppression of the density fluctuations $V_0$.
On the other hand for the case of large fluctuation of the random field 
$\sigma_x\rightarrow \infty$, the $\langle S^x\rangle$-terms in $V_{\rm Rep}$,
\begin{equation}
\Big(2C_1-{4C_1^2 \over V_0}\Big)\langle S^x\rangle^2
-{\beta n \over {1 \over \sigma^2_x}-{\beta n \over 4V_0}}\Big({C_1 \over V_0}
\langle S^x\rangle\Big)^2 \rightarrow 2C_1\langle S^x\rangle^2, \;\;
{\sigma_x\rightarrow \infty},
\end{equation}
and then $S^x$ does not condense whereas $S^y$ does.
Similarly for $T\rightarrow 0$, the last term of $V_{\rm Rep}$ in Eq.(\ref{Vrep})
gives a finite contribution for any nonvanishing $\sigma_x$, and $S^x$ 
does not condense.

The above results seem interesting but they are obtained by the MFT.
For example, the assumption of the constant mean field 
$\langle S^x\rangle$ is not correct for $\sigma_x\rightarrow \infty$.
Therefore more reliable studies are welcome.
The numerical calculations in the subsequent sections give
reliable results and reveal detailed properties of the RFIO.

%%%%%%%%%%%%%%%%%%%%%%%%%%%%%%%%%%%%%%%%%%%%%%%%%%%%%%%
\section{Numerical results}
\subsection{Phase diagrams at low temperature}

%%%%%%%%%%%%%%%%%%%%%%%%%%%%%%%%%%%%%%%%%%%%%%%%%%%%%%%
%FIG.GP1
\begin{figure}[h]
\begin{center}
\includegraphics[width=15cm]{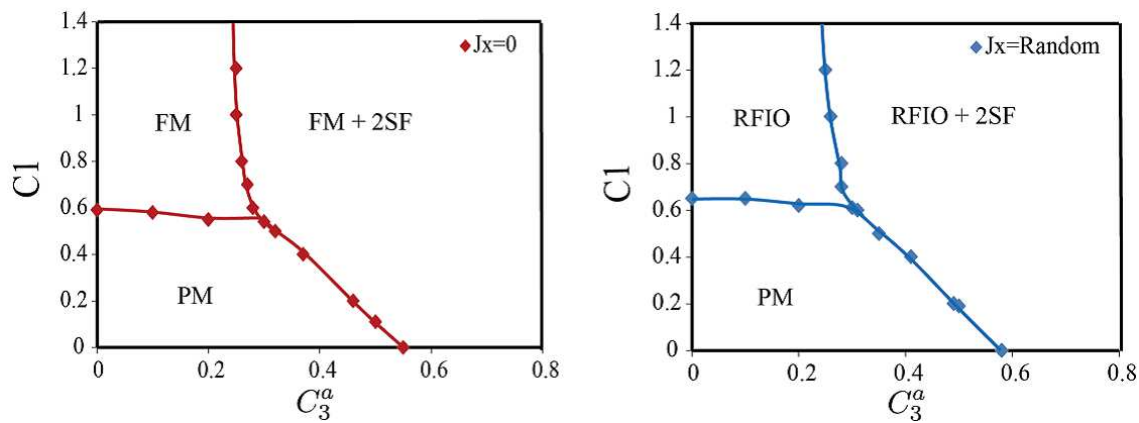}
\vspace{-0.3cm}
\caption{(Color online)
Phase diagrams of the systems $H_{\rm EtJ}$ (left) and 
$H_{\cal T}$ (right) at low temperature.
$c_\tau=2.0, \ C^a_3=C^b_3, \ \sigma_x=0.3$ and $\sigma_y=0$.
RFIO stands for the random-field-induced order.
PM is the state without any orders, FM has the ferromagnetic order
of the pseudo-spin, and SF is the superfluid with the BEC of atom.
All phase transitions are of second order.
}\vspace{-0.5cm}
\label{fig:PD1}
\end{center}
\end{figure}
%%%%%%%%%%%%%%%%%%%%%%%%%%%%%%%%%%%%%%%%%%%%%%%%%%%%%%%%%%%

In this and subsequent sections, we shall show the results obtained by
means of the numerical MC simulations.
Model is defined by Eqs.(\ref{Z2})$\sim$(\ref{Aq}), and the local-update
MC simulation was used for calculation of physical quantities for fixed
random variables $\{\tilde{J}^x_i\} \ (\{\tilde{J}^y_i\})$.
The standard Metropolis algorithm\cite{MC} was used for the local update.
For the local update of the angle variables $\theta_{\alpha i}$,
random variables $\Delta \theta$ used for generating a candidate 
of a new variable $\theta_{\rm new}=\theta_{\rm old}+\Delta \theta$
was chosen in the range $|\Delta \theta|<{\pi \over 3}$.
Typical sweeps for the thermalization is $100000$ and for
the measurement is $(20000)\times$(10 samples).
Typical acceptance ratio is $40\% \sim 50\%$, and 
errors were estimated from 10 samples by the jackknife methods\cite{jack}.

We first show the phase diagram of the system {\em without} the random field,
which was obtained in the previous study\cite{BtJ}.
For the case of $t_a=t_b$ and at $T=0$
\footnote{More precisely in the MC simulations, temperature of the system
$T$ is given as ${k_{\rm B}T \over V_0}={c_\tau \over N_\tau}$,
where $N_\tau$ is the lattice size in the imaginary-time direction.
Then the system at $T\rightarrow 0$ is realized as $N_\tau \rightarrow \infty$.}, 
there are three phases, phase with no long-range order (LRO), 
FM phase and phase of double
BECs, which we often denote 2SF, accompanying the FM order.
See Fig.\ref{fig:PD1}.
In particular in the states with the FM order, the pseudo-spin $(S^x,S^y)$ has
a LRO in an arbitrary direction, i.e., 
$\langle S^y\rangle/\langle S^x\rangle=\tan \theta_s$ with an arbitrary angle $\theta_s$.
This is observed by calculating the correlation functions $G^x_{\rm S}(r)$ and 
$G^y_{\rm S}(r)$ defined by
\begin{equation}
G^{x(y)}_{\rm S}(r)={1 \over L^3}\sum_{r_0}\langle S^{x(y)}(r+r_0)S^{x(y)}(r_0)\rangle,
\;\;\;
G_{\rm S}(r)=G^{x}_{\rm S}(r)+G^{y}_{\rm S}(r),
\label{Gs}
\end{equation}
where $L$ is the linear size of the 3D lattice, and
sites $r_0$ and $r_0+r$ are located in the same spatial 2D lattice,
i.e., $G^{x(y)}_{\rm S}(r)$ is an equal-time correlator. 
We put the lattice size of the imaginary-time direction 
$N_\tau=\mbox{the lattice size of the spatial direction }L$.
The angle $\theta$ takes various values depending on initial configurations and 
random variables used local updates in the MC simulations.
This result comes from the U(1) symmetry of the pseudo-spin rotation in the
system of the action $A_{{\rm L}\tau}+A_{\rm L}$.

%%%%%%%%%%%%%%%%%%%%%%%%%%%%%%%%%%%%%%%%%%%%%%%%%%%%%%%%%%%
%FIG.GP1
\begin{figure}[h]
\begin{center}
\includegraphics[width=5cm]{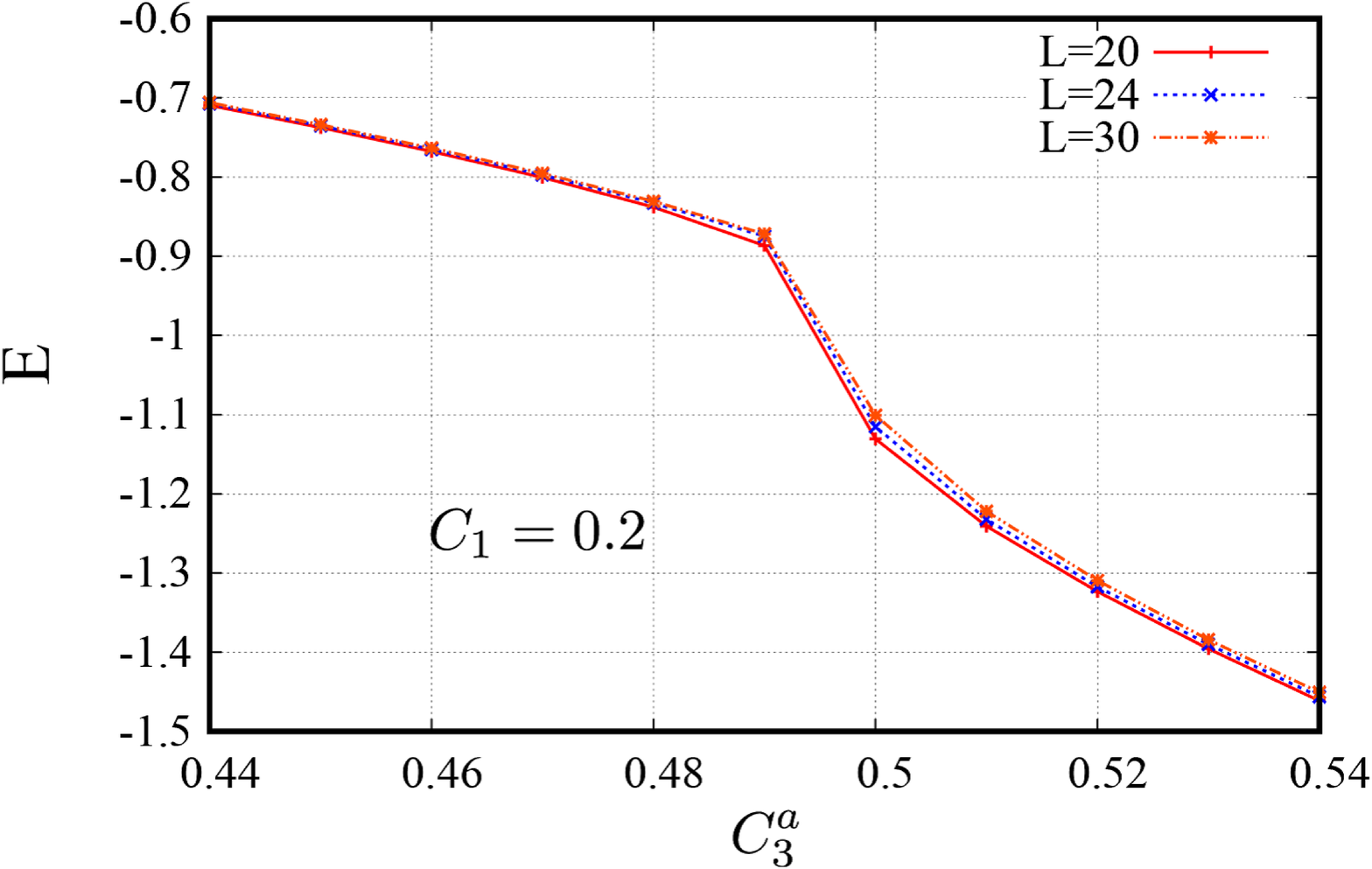}
\includegraphics[width=5cm]{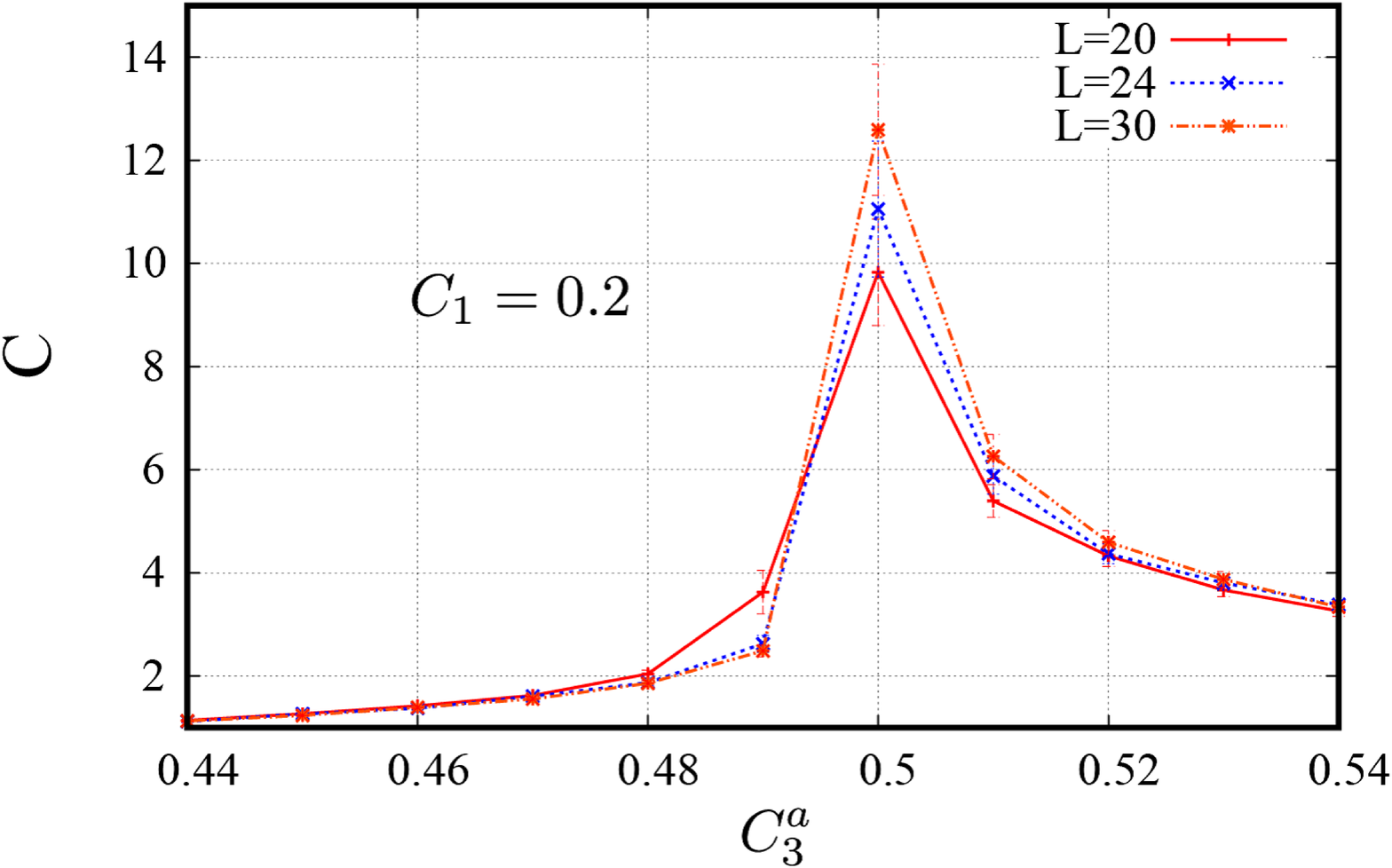}
\includegraphics[width=5cm]{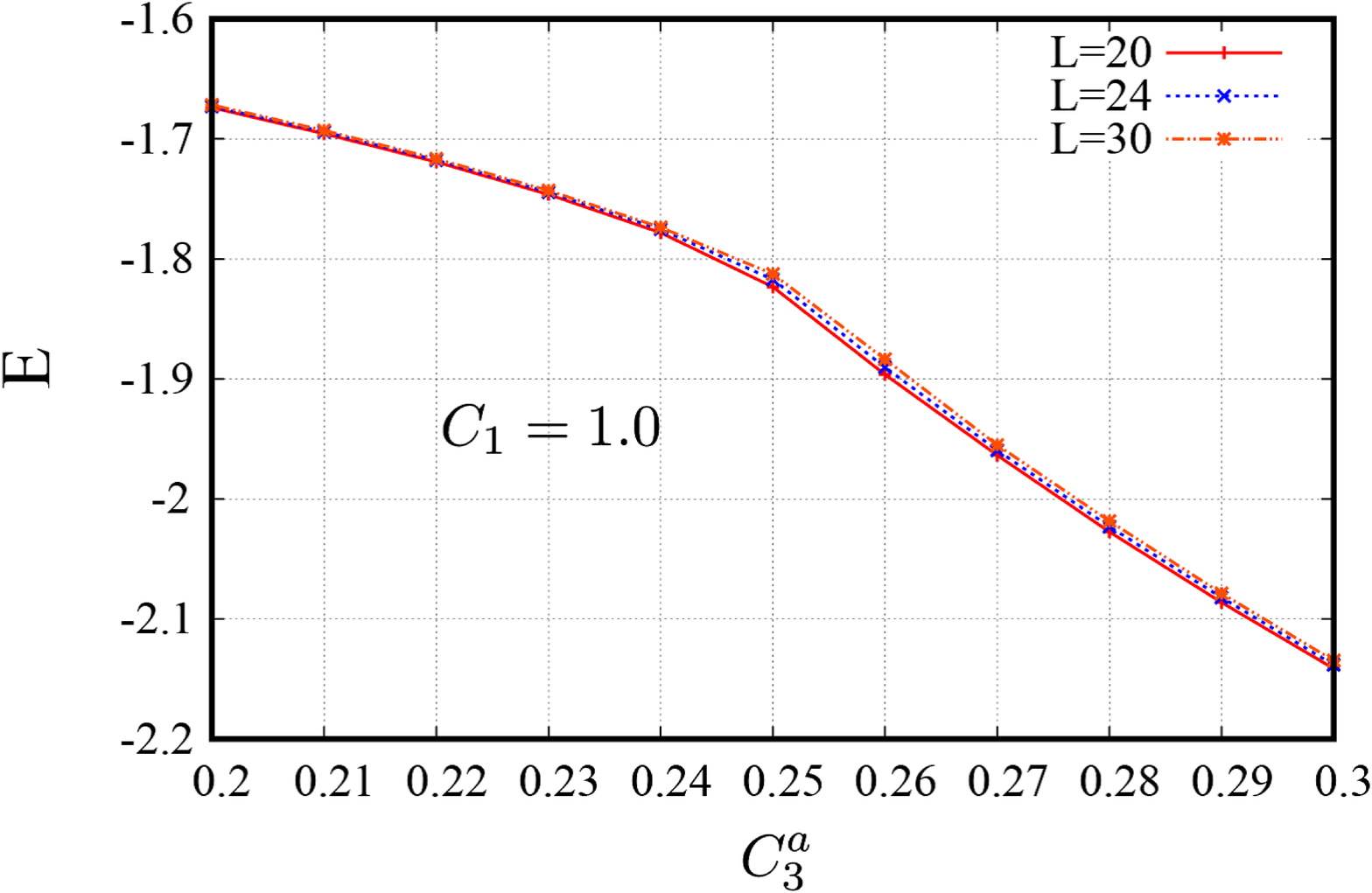}
\includegraphics[width=5cm]{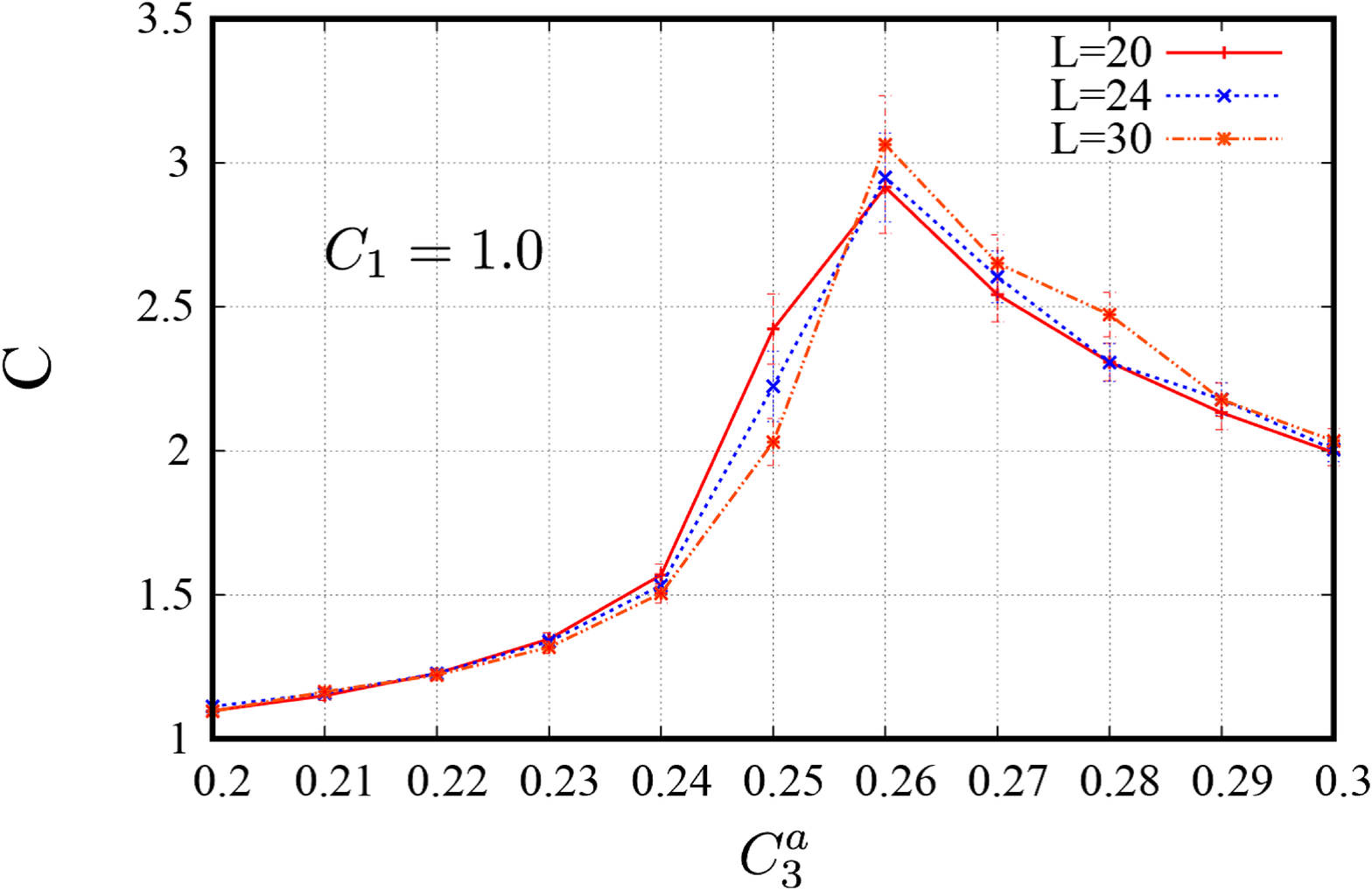}
\includegraphics[width=5cm]{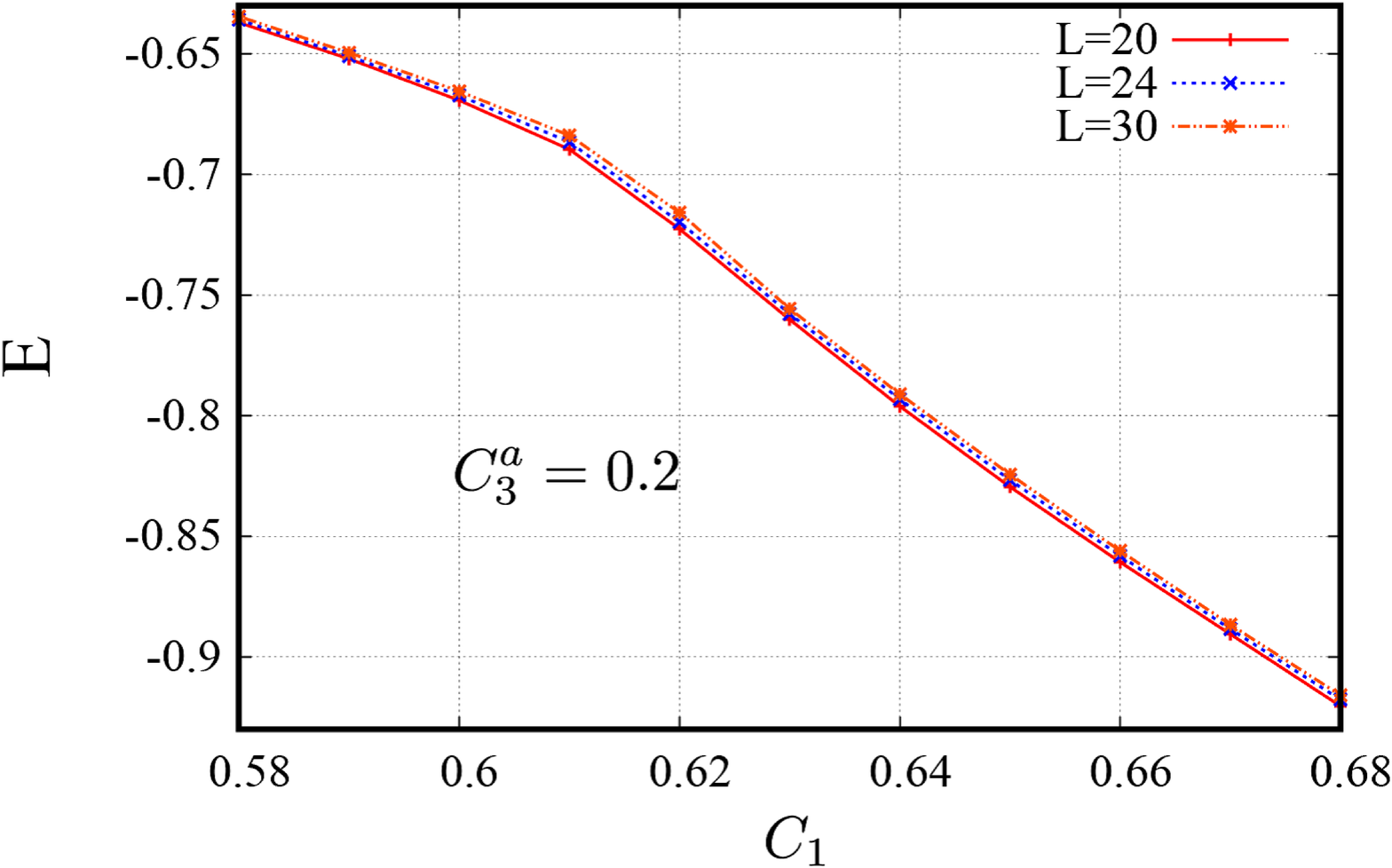}
\includegraphics[width=5cm]{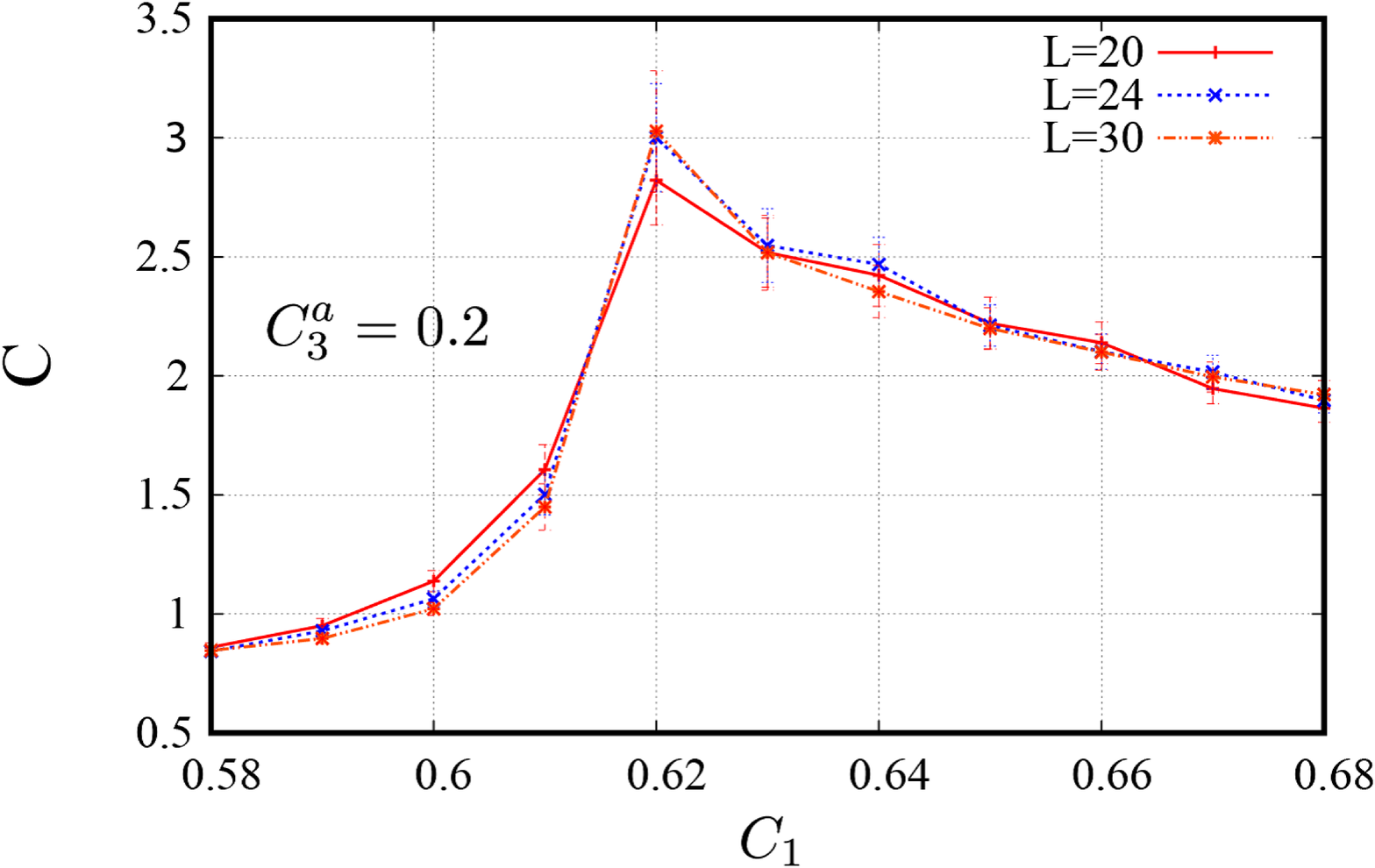}
\vspace{-0.3cm}
\caption{(Color online)
Internal energy $E$ and specific heat $C$ for various system sizes.
Results indicate the existence of second-oder phase transitions.
$C^a_3=C^b_3$ and $\sigma_x=0.3$.
}\vspace{-0.5cm}
\label{fig:EC1}
\end{center}
\end{figure}
%%%%%%%%%%%%%%%%%%%%%%%%%%%%%%%%%%%%%%%%%%%%%%%%%%%%%%%%%%%
%%%%%%%%%%%%%%%%%%%%%%%%%%%%%%%%%%%%%%%%%%%%%%%%%%%%%%%%%%%
%FIG.GP1
\begin{figure}[h]
\begin{center}
\includegraphics[width=10cm]{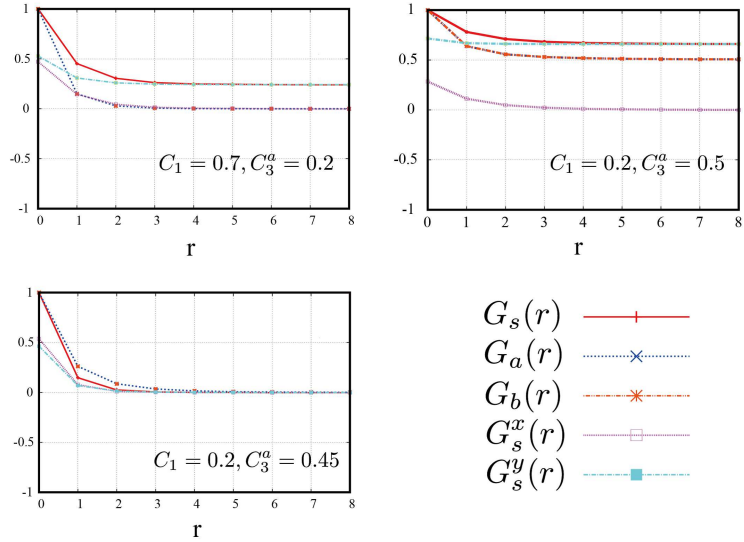}
\vspace{-0.3cm}
\caption{(Color online)
Correlation functions, which are used to identify the various phases.
$G_{\rm S}(r), \ G_a(r), \ G_b(r), \ G^x_{\rm S}(r)$, and 
$G^y_{\rm S}(r)$.
$C^a_3=C^b_3$ and $\sigma_x=0.3$.
}\vspace{-0.5cm}
\label{fig:CF1}
\end{center}
\end{figure}
%%%%%%%%%%%%%%%%%%%%%%%%%%%%%%%%%%%%%%%%%%%%%%%%%%%%%%%%%%%

In the present paper, we shall study the system in the random fields $A_{\rm q}$
in Eq.(\ref{Aq}).
We first consider the case $\tilde{J}^y_i=0$.
In this single-component external system, the U(1) spin symmetry is
reduced to the $Z_2$ symmetry of the Ising type $(S^x,S^y)\rightarrow -(S^x,S^y)$.
This fact implies that there might exist a preferred direction in the pseudo-spin order.
This is actually the case as we show shortly.

In the numerical studies, we first determine the random variables $\{\tilde{J}^x_i\}$
according to the distribution $P(\tilde{J}_x)$ in Eq.(\ref{PJ}).
To this end, we used the box-Muller methods.
Then the MC simulation is carried out for the system with the fixed $\{\tilde{J}^x_i\}$
by the local update of $\omega_{\alpha r}$ and $\lambda_r$.
Final physical quantities are obtained by averaging calculated quantities for
each sample over $5\sim 10$ $\{\tilde{J}^x_i\}$ samples.
Phase boundary is determined by calculating the internal energy $E$ and 
the specific heat $C$, which are defined as 
\begin{eqnarray}
E=\langle A_{\rm Lxy} \rangle/L^3, \;\;\;
C=\langle (A_{\rm Lxy}-E)^2 \rangle/L^3,
\label{EC}
\end{eqnarray}
where the mean value $\langle \cdot \rangle$ includes the average over samples
of the random fields as we explained above.
In order to identify physical properties of each phase, we also calculate 
the boson correlation functions besides the pseudo-spin one in Eq.(\ref{Gs}),
\begin{eqnarray}
&& G_{a}(r)={1 \over L^3}\sum_{r_0} \langle 
e^{i\theta_{ar_0}}e^{-i\theta_{a,r_0+r}}\rangle, \nonumber \\
&&G_{b}(r)={1 \over L^3}\sum_{r_0} \langle 
e^{i\theta_{br_0}}e^{-i\theta_{b,r_0+r}}\rangle,
\label{CF1}
\end{eqnarray}
where, as in $G^{x(y)}_{\rm S}(r)$, 
sites $r_0$ and $r_0+r$ are located in the same spatial 2D lattice.

\begin{figure}[h]
\begin{center}
\includegraphics[width=8cm]{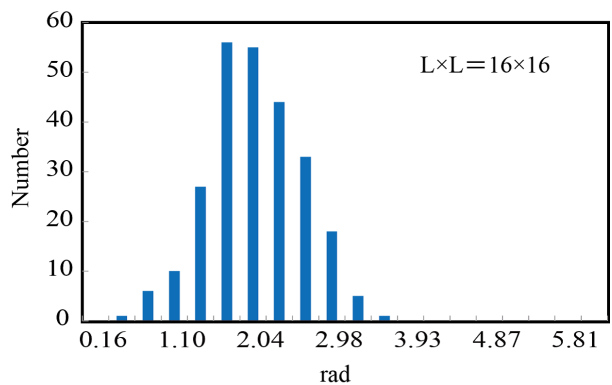}
\vspace{-0.3cm}
\caption{(Color online)
Histogram of the relative phase of the condensation of the $a$- and $b$-atoms
at constant $\tau$.
The results clearly indicate the relative phase 
$\theta_s=\theta_a-\theta_b\simeq {\pi \over 2}$.
$C_1=2.0, \ C^a_3=C^b_3=0.2, \ c_\tau=2.0$ and $\sigma_x=0.3$.
}\vspace{-0.5cm}
\label{fig:phaseH1}
\end{center}
\end{figure}

In Fig.\ref{fig:PD1}, we exhibit the obtained phase diagram for $\sigma_x=0.3$ and 
$\sigma_y=0$, which corresponds to $\tilde{J}^y_i=0$.
Typical behaviors of $E$ and $C$ of various system sizes, which are used for
identification of the phase boundaries, are also shown in Fig.\ref{fig:EC1}.
The calculations indicate that all phase transitions are of second order.
The locations of the phase boundaries are almost the same with the ones of the
original bosonic t-J model.
However the FM order is replaced by the RFIO, as the pseudo-spin 
correlation function in Fig.\ref{fig:CF1} indicates that only the $y$-component of
the pseudo-spin has a LRO and the correlation of the $x$-component
vanishes quite rapidly as a function of $r$.
This result means that the relative phase of the condensations of
the $a$- and $b$-boson operators
has definite values $\theta_s\simeq\pm {\pi \over 2}$.
To verify this, we measured the relative phase at each site and the
result is shown in Fig.\ref{fig:phaseH1}.
The magnetization in the phase of RFIO is slightly smaller than that in the
original t-J model.
The above observation is in good agreement with the previous studies of the related
systems such as the classical XY spin model and two component BEC of 
the cold atoms\cite{RFIO1,RFIO2}. 

%%%%%%%%%%%%%%%%%%%%%%%%%%%%%%%%%%%%%%%%%%%%%%%%%%%%%%%%%%%
%FIG.GP1
%%%%%%%%%%%%%%%%%%%%%%%%%%%%%%%%%%%%%%%%%%%%%%%%%%%%%%%%%%%
%FIG.GP1
\begin{figure}[h]
\begin{center}
\includegraphics[width=7cm]{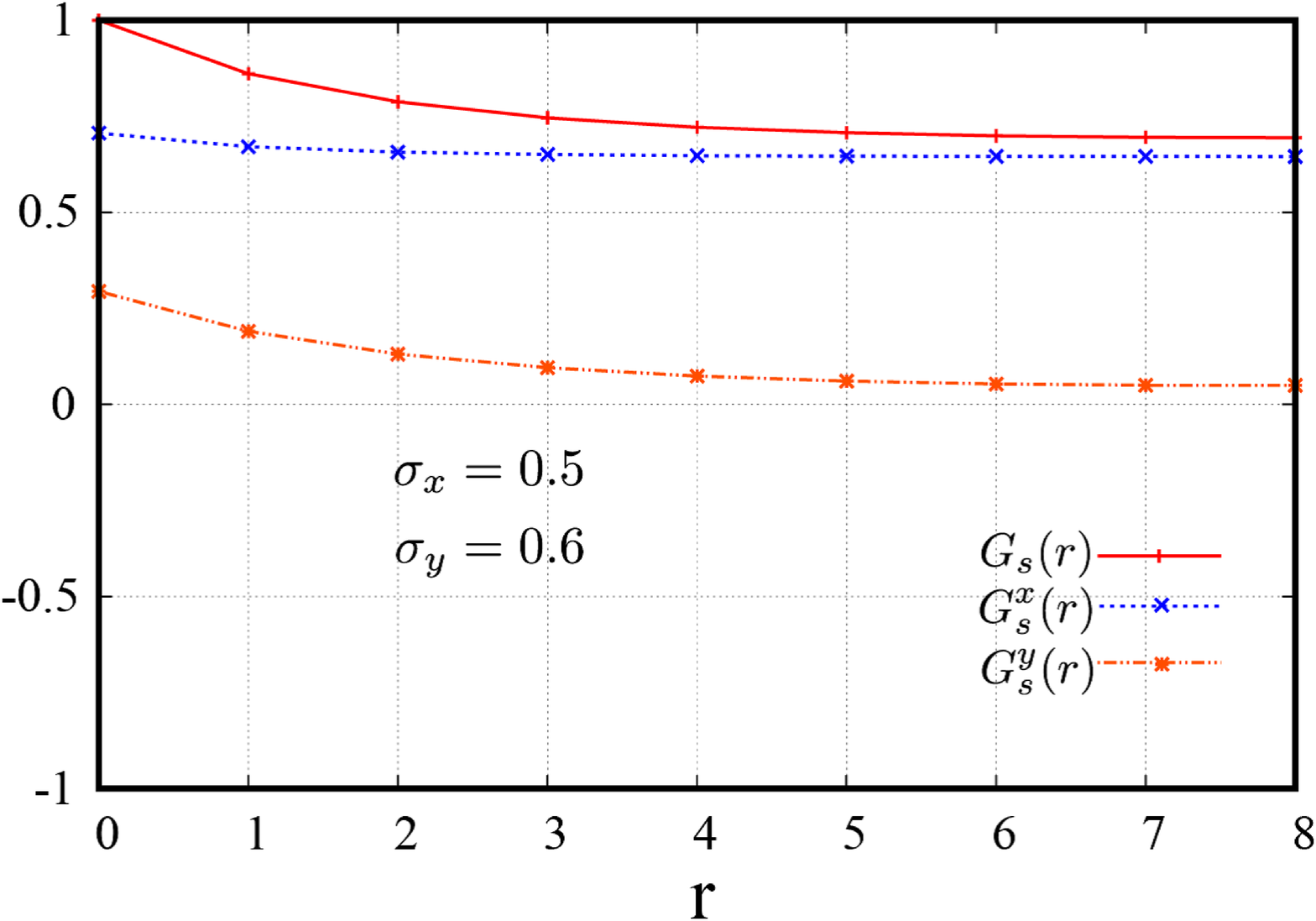}
\includegraphics[width=7cm]{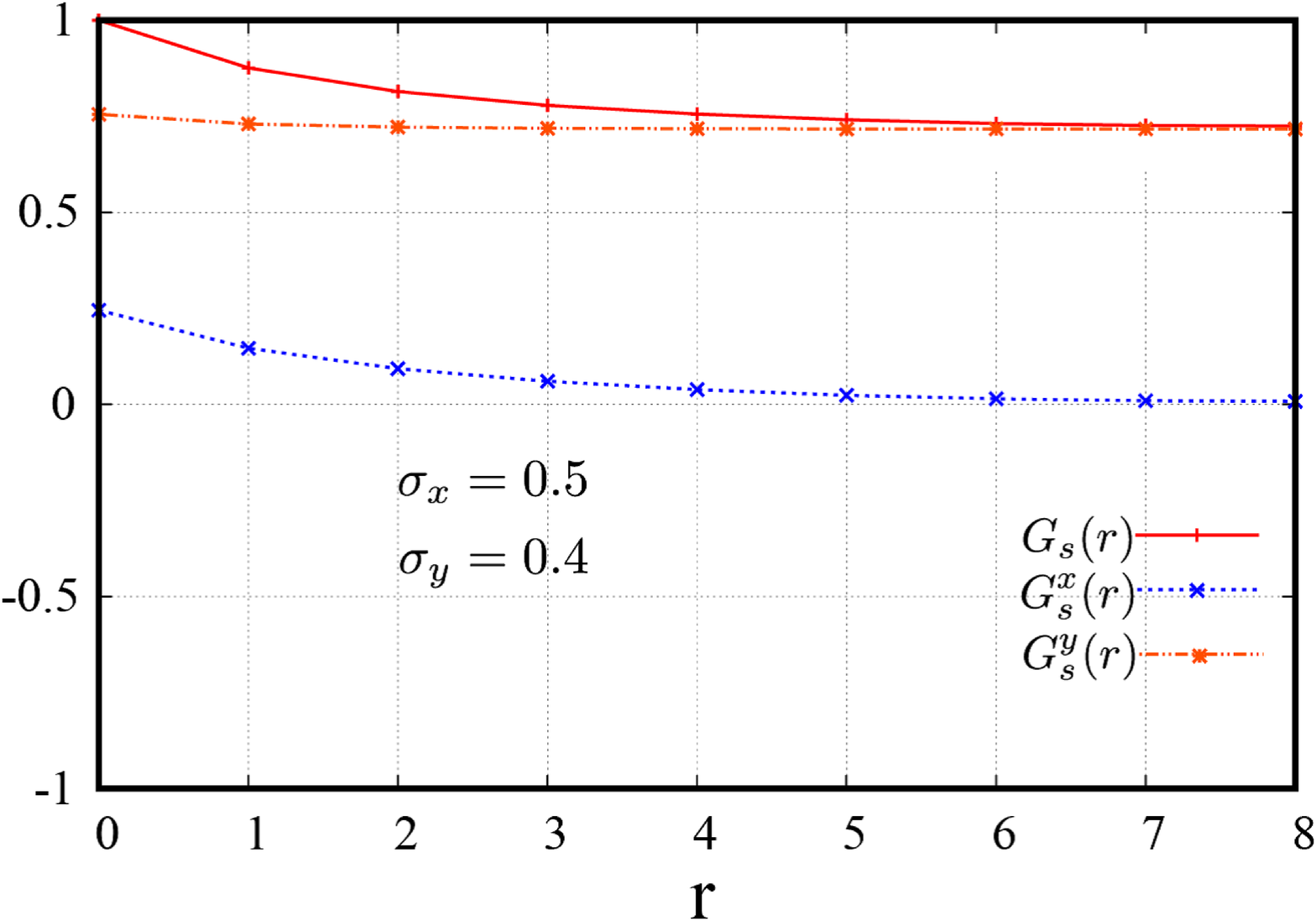}
\vspace{-0.3cm}
\caption{(Color online)
Pseudo-spin correlation function for various $\sigma_x$ and $\sigma_y$.
$G_{\rm S}(r), \ G_{\rm S}^x(r)$, and $G_{\rm S}^y(r)$.
Orientation of the magnetization of the pseudo-spin is determined by 
relative magnitude of $\sigma_x$ and $\sigma_y$.
}\vspace{-0.5cm}
\label{fig:CF2}
\end{center}
\end{figure}
%%%%%%%%%%%%%%%%%%%%%%%%%%%%%%%%%%%%%%%%%%%%%%%%%%%%%%%%%%%

It is interesting to see how the phase diagram is changed when both components
of the random field are turned on.
It is not so difficult to show that the U(1) symmetry is restored
for the case $\sigma_x=\sigma_y$, and then the relative phase 
$\theta_s=\theta_a-\theta_b$ takes an
arbitrary value.
From this observation, one may expect that the range of $\theta$ is 
expanded for nonvanishing
$\sigma_x$ and $\sigma_y$ depending on the ratio $\sigma_x/\sigma_y$.
However this is not the case.
We studied the cases with various value of $\sigma_x/\sigma_y$, and
found that $\theta_s$ takes $\pm {\pi \over 2} \ (0 \ \mbox{or} \ \pi)$
for $\sigma_x/\sigma_y>1 \ (<1)$.
For a typical example, see Fig.\ref{fig:CF2}.

Let us perform the ``Gedanken experiment" in which 
the parameter $V_0$ is varied with the other parameters fixed.
For larger $V_0$, fluctuations of the densities of atoms in each site are suppressed 
and then fluctuations of the phase degrees of freedom of the boson operators are enhanced.
In fact from the action in Eqs.(\ref{Atau}) and (\ref{parameters}), 
it is seen that the phases $\omega_{\alpha r}$ vary rapidly in the $\tau$-direction for small $c_\tau$, even though their order are 
generated in the spatial direction for a sufficiently large hopping amplitude and
a spin-exchange coupling.

%%%%%%%%%%%%%%%%%%%%%%%%%%%%%%%%%%%%%%%%%%%%%%%%%%%%%%%%%%%
%FIG.GP1
\begin{figure}[h]\vspace{-1.0cm}
\begin{center}
\includegraphics[width=10cm]{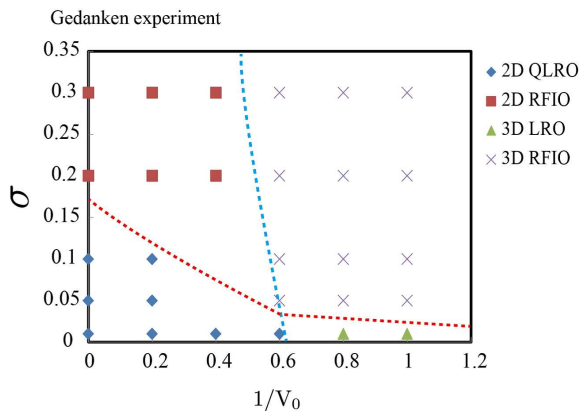}
\vspace{-0.3cm}
\caption{(Color online)
Phase diagram in the $({1 \over V_0}-\sigma_x)$ plain.
(We put $\Delta \tau=1$.)
$C_1=2.0$ and $C^a_3=C^b_3=0.2$.
The dotted lines denote crossover lines and properties of each ``phase"
are identified calculating the correlation functions.
}\vspace{-0.5cm}
\label{fig:V0PD}
\end{center}
\end{figure}
%%%%%%%%%%%%%%%%%%%%%%%%%%%%%%%%%%%%%%%%%%%%%%%%%%%%%%%%%%%

We show the results of the numerical study in Fig.\ref{fig:V0PD}.
Phase diagram is given in the $({1 \over V_0}-\sigma_x)$ plane for
$C_1=2.0$ and $C^a_3=C^b_3=0.2$.
In the 3D region of smaller $V_0$, phase transition from the 3D XY-spin ordered 
state to the 3D RFIO takes place as $\sigma_x$ is increased.
Both the states have the own LROs.
On the other hand for larger $V_0$, the system has a quasi-LRO for smaller
$\sigma_x$, and the state turns to that of the RFIO with the genuine $Z_2$
LRO as $\sigma_x$ is increased.
In the limit ${1 \over V_0}\rightarrow 0$, the system can be regarded as 
a classical 2D system.
Therefore the study of this limit reproduces the result of the previous study
on the classical XY model in 2D with the random external field\cite{RFIO1}.
There the RFIO forms simply as properties of the lowest energy state.

%%%%%%%%%%%%%%%%%%%%%%%%%%%%%%%%%%%%%%%%%%%%%%%
\subsection{Robustness of a finite RFIO state}

%%%%%%%%%%%%%%%%%%%%%%%%%%%%%%%%%%%%%%%%%%%%%%%%%%%%%%%%%%%
%FIG.GP1
\begin{figure}[h]
\begin{center}
\includegraphics[width=10cm]{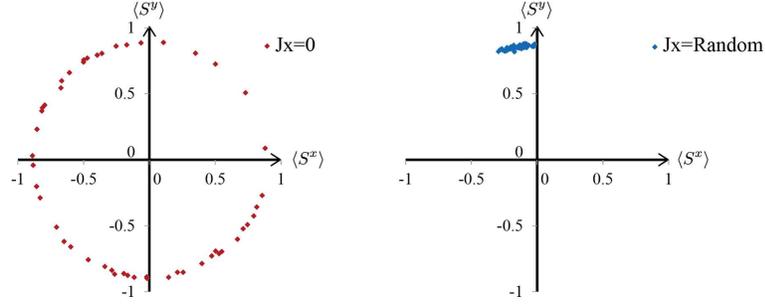}
%\vspace{-0.3cm}
\caption{(Color online)
Behavior of the average of the pseudo-spin 
$(\langle S^x\rangle, \langle S^y\rangle )$ for fixed $\tau$ under the MC local update.
In the system without the random external field (left), the pseudo-spin
fluctuates strongly under the MC update, whereas the random 
external field stabilizes the orientation of the pseudo-spin to the Ising type (right).
The number of the total sweep is $5\times 10^5$.
Data points are plotted for every $10^4$ sweeps.
}\vspace{-0.5cm}
\label{fig:spinorder}
\end{center}
\end{figure}
%%%%%%%%%%%%%%%%%%%%%%%%%%%%%%%%%%%%%%%%%%%%%%%%%%%%%%%%%%%
%%%%%%%%%%%%%%%%%%%%%%%%%%%%%%%%%%%%%%%%%%%%%%%%%%%%%%%%%%%
%FIG.GP1
\begin{figure}[h]
\begin{center}
\includegraphics[width=5cm]{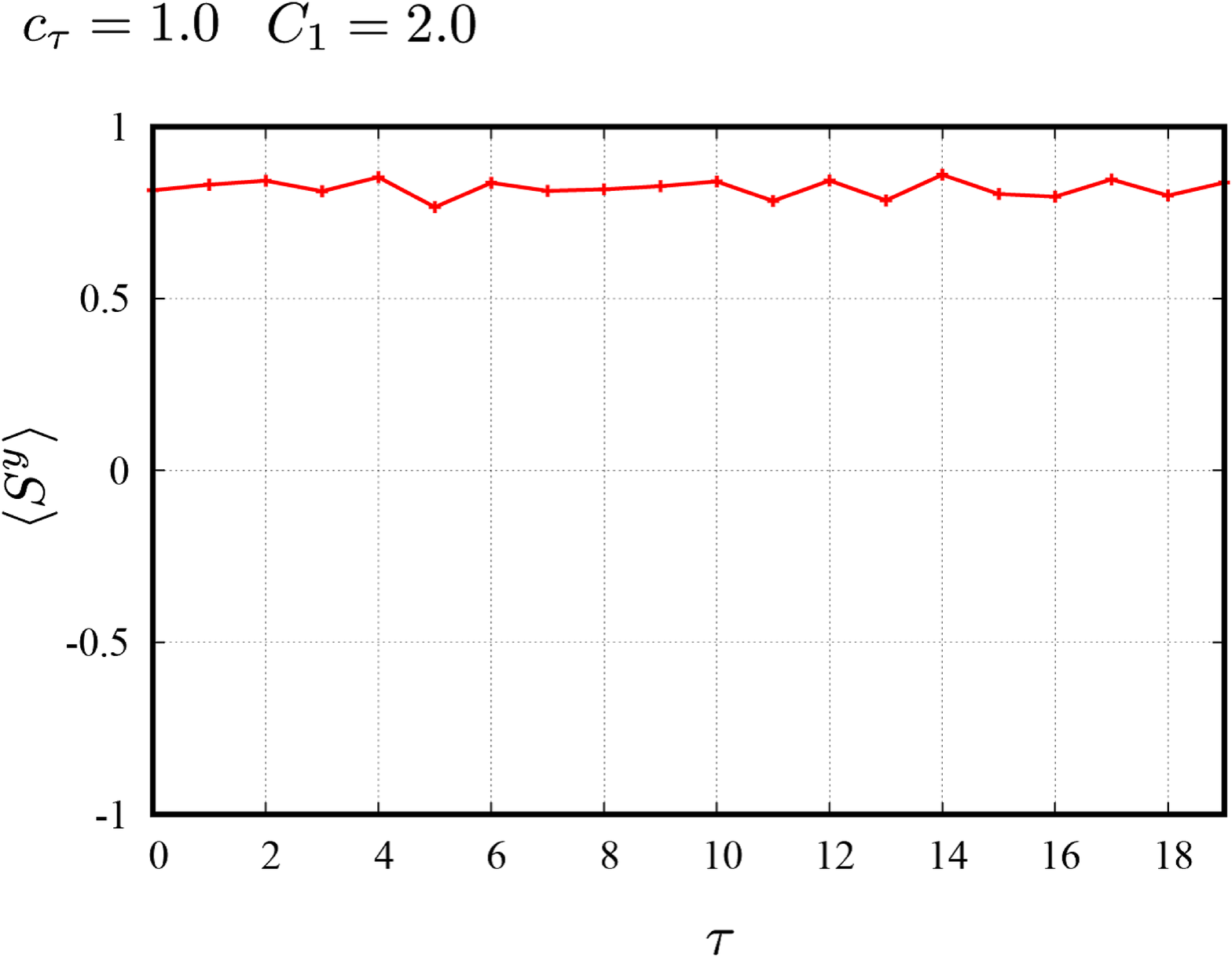} \hspace{0.5cm}
\includegraphics[width=5cm]{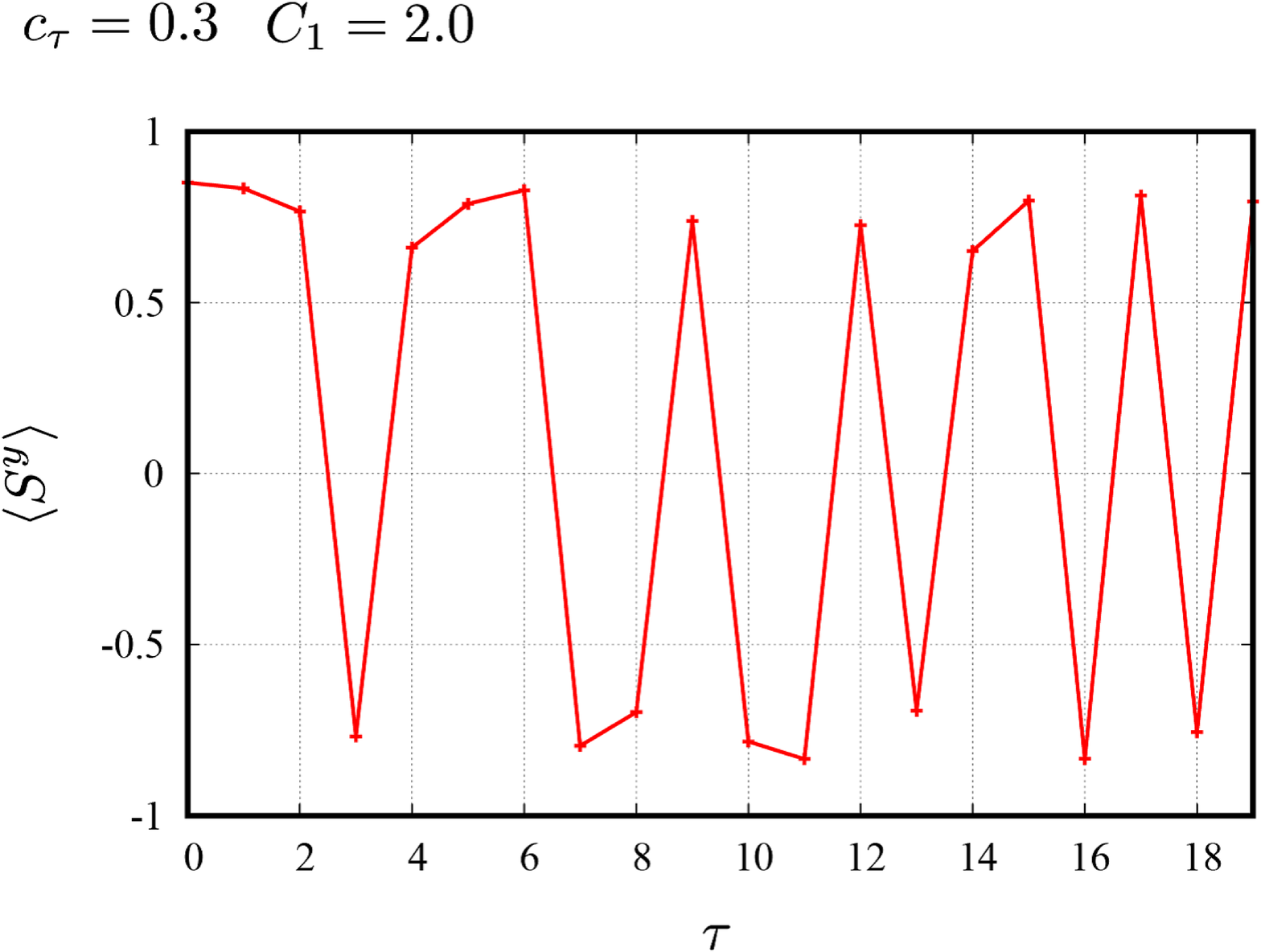}
%\vspace{-0.3cm}
\caption{(Color online)
Behavior of the 2D average of the pseudo-spin 
$\langle S^y\rangle$ as a function of the imaginary-time $\tau$. 
For $c_\tau=2.0$, $\langle S^y\rangle$ is almost constant and behaves
as a classical spin.
On the other hand for $c_\tau=0.3$, the direction of $\langle S^y\rangle$
changes as a result of quantum fluctuations.
}\vspace{-0.5cm}
\label{fig:spintau}
\end{center}
\end{figure}
%%%%%%%%%%%%%%%%%%%%%%%%%%%%%%%%%%%%%%%%%%%%%%%%%%%%%%%%%%%

The phase diagram obtained in the previous section, Fig.\ref{fig:V0PD},
shows that the state of the quasi-LRO changes to the state with 
the genuine Ising-type RFIO as $\sigma_x$ increases for a large $V_0$.
This result indicates the robustness of the RFIO state.
In order to verify this fact, we study how finite-size systems
of the quasi-LRO and also the RFIO change under the updates of the MC simulations.
Result of $(\langle S^x \rangle, \langle S^y \rangle)$ for {\em fixed} $\tau$
is shown in Fig.\ref{fig:spinorder}.
Parameters are $c_\tau=2.0(={1 \over V_0\Delta \tau})$, $C_1=2.0$ 
and $C^a_3=C^b_3=0.2$, and system size is $10^3$.
For the case of the RFIO, $\sigma_x=0.3$.
Number of the total MC sweep is $5\times 10^5$.
Data are plotted for every $10^4$ sweeps.
Similar result to that of the RFIO in Fig.\ref{fig:spinorder} is also obtained in the
2D RFIO state in Fig.\ref{fig:V0PD}.

From Fig.\ref{fig:spinorder},
it is obvious that, in the ordinary system without the random external field,
the average of the pseudo-spin (magnetization) is nonvanishing but 
unstable under the MC update,
i.e., the orientation of the magnetization fluctuates strongly because of the
finiteness (smallness) of the system.
On the other hand in the random case, the average is quite stable and stays
$\langle S^y_i\rangle=\pm 1, \ \langle S^x_i \rangle=0$.
Then one may wonder that the spin behaves as a classical spin and 
lost its quantum properties.
In order to study it, we measured behavior of 
$(\langle S^x \rangle, \langle S^y \rangle)$ as a function of  $\tau$.
See Fig.\ref{fig:spintau}.
The results in Fig.\ref{fig:spintau} indicate the following fact.
For a small $V_0$ (i.e., large $c_\tau$), 
the phase degrees of freedom of the boson operators
$a_i$ and $b_i$ have small quantum fluctuations as their the boson densities
fluctuate rather largely.
Then, the spin behaves as a classical spin.
On the other hand  for a large $V_0$ (small $c_\tau$), 
the spin behaves as a quantum spin
and a superposition of the $\uparrow$ spin and $\downarrow$ spin is
possible.
This result indicates that the 2D RFIO state can be used as a quantum qubit
in the quantum information device.
%%%%%%%%%%%%%%%%%%%%%%%%%%%%%%%%%%%%%%%%%%%%%%%%%%%%%%%%%%%
%FIG.Entangle
\begin{figure}[h]
\begin{center}
\includegraphics[width=8cm]{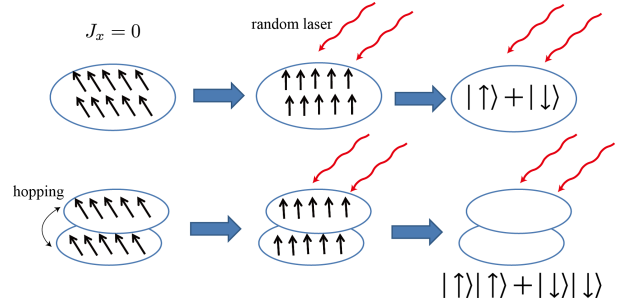} 
%\vspace{-0.3cm}
\caption{(Color online)
Methods of making a superposed state of single qubit and an entangled state
of two qubits.
In the vanishing random-external field, spin is unstable as a result of the quantum
fluctuations.
By applying a random field suddenly to that state, a superposed state
is expected to form.
}\vspace{-0.5cm}
\label{fig:entangle}
\end{center}
\end{figure}
%%%%%%%%%%%%%%%%%%%%%%%%%%%%%%%%%%%%%%%%%%%%%%%%%%%%%%%%%%%
In Fig.\ref{fig:entangle}, we show a method to make a superposed state 
$|\uparrow\rangle+|\downarrow\rangle$ and also an entangled state
of two qubits.
Study on quantum superpositions of macroscopically distinct states
has the long history\cite{superpose}.
The above study indicates the possibility that a mesoscopic RFIO state
is a candidate for the quantum mesoscopic superposed state.

%%%%%%%%%%%%%%%%%%%%%%%%%%%%%%%%%%%%%%%%%%%%%%%
\subsection{Finite temperature phase diagram}

%%%%%%%%%%%%%%%%%%%%%%%%%%%%%%%%%%%%%%%%%%%%%%%%%%%%%%%%%%%
%FIG.GP1
\begin{figure}[h]
\begin{center}
\includegraphics[width=10cm]{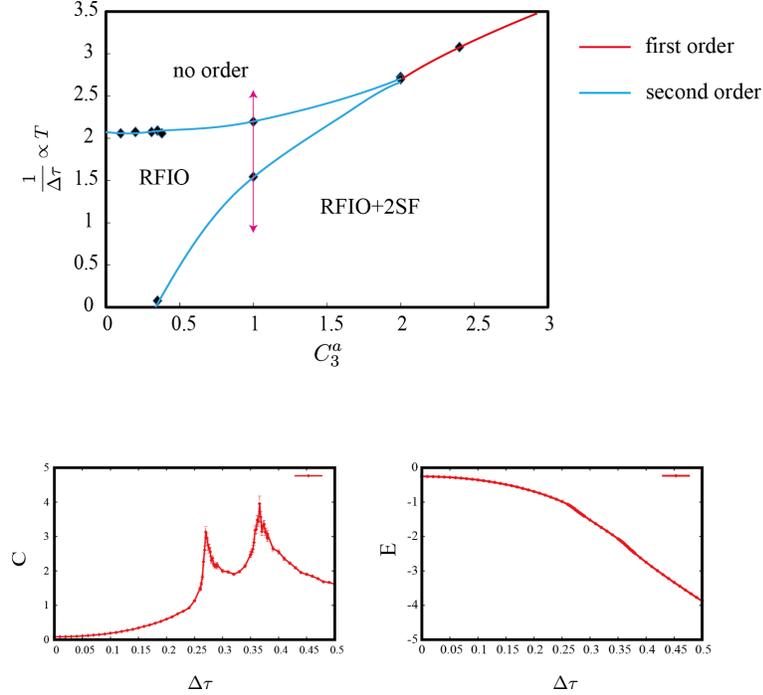}
%\vspace{-0.3cm}
\caption{(Color online)
Phase diagram in the $(T-C^a_3)$ plain, and the internal energy $E$ and 
specific heat $C$ as a function of $\Delta\tau$.
$C_1=3.0, \ C^a_3=C^b_3$ and $\sigma_x=0.3$.
The blue lines indicate second-order phase transition lines, whereas 
the red one denotes a first-order phase transition line.
}\vspace{-0.5cm}
\label{fig:FTPD}
\end{center}
\end{figure}
%%%%%%%%%%%%%%%%%%%%%%%%%%%%%%%%%%%%%%%%%%%%%%%%%%%%%%%%%%%

In this subsection, we shall study finite-$T$ phase diagram of the random
system.
In particular, we are interested in how the states with the RFIO evolve 
as $T$ is increased.
This study is closely related with the stability of the RFIO states
investigated in the previous subsection..
%If robustness of the states is shown, a cold atomic system with the RFIO
%can be used as qubits for quantum calculation.

In the present MC simulation, the temperature is given as 
$k_{\rm B}T=1/(N_\tau\Delta\tau)$.
Then, system at low $T$ is realized for sufficiently large $N_\tau$.
Temperature of the system is increased by decreasing $\Delta\tau$
for fixed $N_\tau$, and therefore
the parameters in the action vary as indicated in Eqs.(\ref{parameters}).
It is obvious that the original 3D system tends to be quasi-1D as 
$\Delta\tau\rightarrow\mbox{small}$, and then the LROs disappear,
which is nothing but a finite-$T$ phase transition.

It is interesting how the ordered states evolve as $T$ is increased.
In order to identify the finite-$T$ phase diagram, we measured 
the internal energy $E$ and the specific heat $C$ as the investigation
of the quantum phase transition in the previous section.
We also calculated the various correlation functions to identify each
phase transition.
We show the obtained results and the phase diagram in Fig.\ref{fig:FTPD}. 
In Fig.\ref{fig:FTPD}, for the states of the 2SF for moderate hopping amplitude
$C^a_3=C^b_3$,
the specific heat $C$ exhibits two sharp peaks that indicate a second-order
phase transition.
The correlation functions show that the 2SF state first loses the properties 
of the BECs and then the pseudo-spin LRO as $T$ is increased.
On the other hand for deep 2SF state, a first-oder phase transition takes
place from the 2SF to the disordered state directly.
This disordered state should be distinguished from the PM state in the low $T$
phase diagram.
The latter appears as a result of the competition between $V_0$-term
and the hopping, whereas the present one comes from the effect of the thermal
fluctuations.

%%%%%%%%%%%%%%%%%%%%%%%%%%%%%%%%%%%%%%%%%%%%%%%%%%%%%%%%%

\section{Topological excitations in RFIO state}

It is interesting to compare topological excitations in the genuine
FM state and RFIO state.
There are two topological objects that play an important role
near phase boundary;
\begin{enumerate}
\item vortices of the $a$ and $b$-atoms and their bound state
\item domain wall of the relative phase of the BECs of  the $a$ and $b$-atoms
\end{enumerate}

It is well known that the $a$ and $b$-vortices proliferate in the PM state,
whereas, in the FM state, the spatial overlap of these two kind of vortex increases
as a result of the coherent condensation of the spin operator $S^x_i$
and/or $S^y_i$. 
Furthermore in a constant external magnetic field $\vec{h}$, the Zeeman coupling 
$\vec{h}\cdot\vec{S}_i$ generates a linear-potential between the
$a$ and $b$-vortices and ``confinement of vortices" takes place.
For example $\vec{h}=(h,0)$, the Zeeman coupling is given as
$hS^x_i=h\cos (\theta_{ai}-\theta_{bi})$,
where $\theta_{ai} \ (\theta_{bi})$ is the phase of the $a (b)$-atom,
and then for a configuration of vortex pair, an extra energy is generated 
proportional to $h\cdot$(distance between two vortices in a pair).
In two-gap superconductors, a mixing  of the two Cooper
pairs gives a similar effect to the Zeeman coupling in the FM, 
and therefore it generates a confinement of vortex\cite{confvortex}.

%%%%%%%%%%%%%%%%%%%%%%%%%%%%%%%%%%%%%%%%%%%%%%%%%%%%%%%%%%%
%FIG.GP1
\begin{figure}[h]
\begin{center}
\includegraphics[width=12cm]{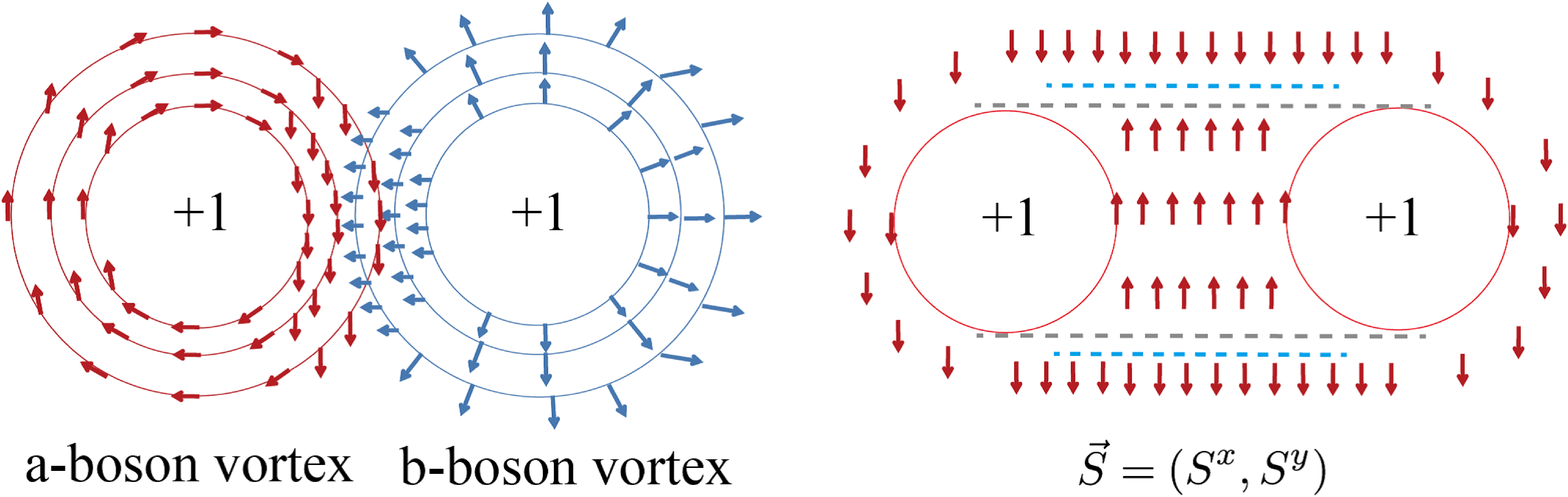} \hspace{0.5cm}
\vspace{-0.3cm}
\caption{(Color online)
(Left) Typical configuration of a $a$- and $b$-atom vortex pair.
Arrows indicate phases of BECs, $\theta_{ai}$ and $\theta_{bi}$.
(Right) Spin configuration $\vec{S}_i=(S^x_i,S^y_i)$ corresponding to
the vortex pair. 
There exist two brick domain walls.
}\vspace{-0.5cm}
\label{fig:vortexp}
\end{center}
\end{figure}
%%%%%%%%%%%%%%%%%%%%%%%%%%%%%%%%%%%%%%%%%%%%%%%%%%%%%%%%%%%

In the system in a random external magnetic field, 
the Zeeman coupling such as $\tilde{J}^x_i\cos (\theta_{ai}-\theta_{bi})$
shows up.
As seen in the previous section, configurations like 
$(\theta_{ai}-\theta_{bi})\sim \pm{\pi \over 2}$ dominates because of
the random Zeeman coupling.
Therefore it is expected that the interaction between a pair of $a$-vortex and 
$b$-vortex in the RFIO
is quantitatively different from that in a constant magnetic field, i.e., a constant Rabi oscillation that prefers configurations with $(\theta_{ai}-\theta_{bi})\sim 0$.
We investigate this problem in this section.
Expected configuration of a vortex pair in the RFIO state is shown in Fig.\ref{fig:vortexp}.

%%%%%%%%%%%%%%%%%%%%%%%%%%%%%%%%%%%%%%%%%%%%%%%%%%%%%%%%%%%
%FIG.GP1
\begin{figure}[h]
\begin{center}
\includegraphics[width=5cm]{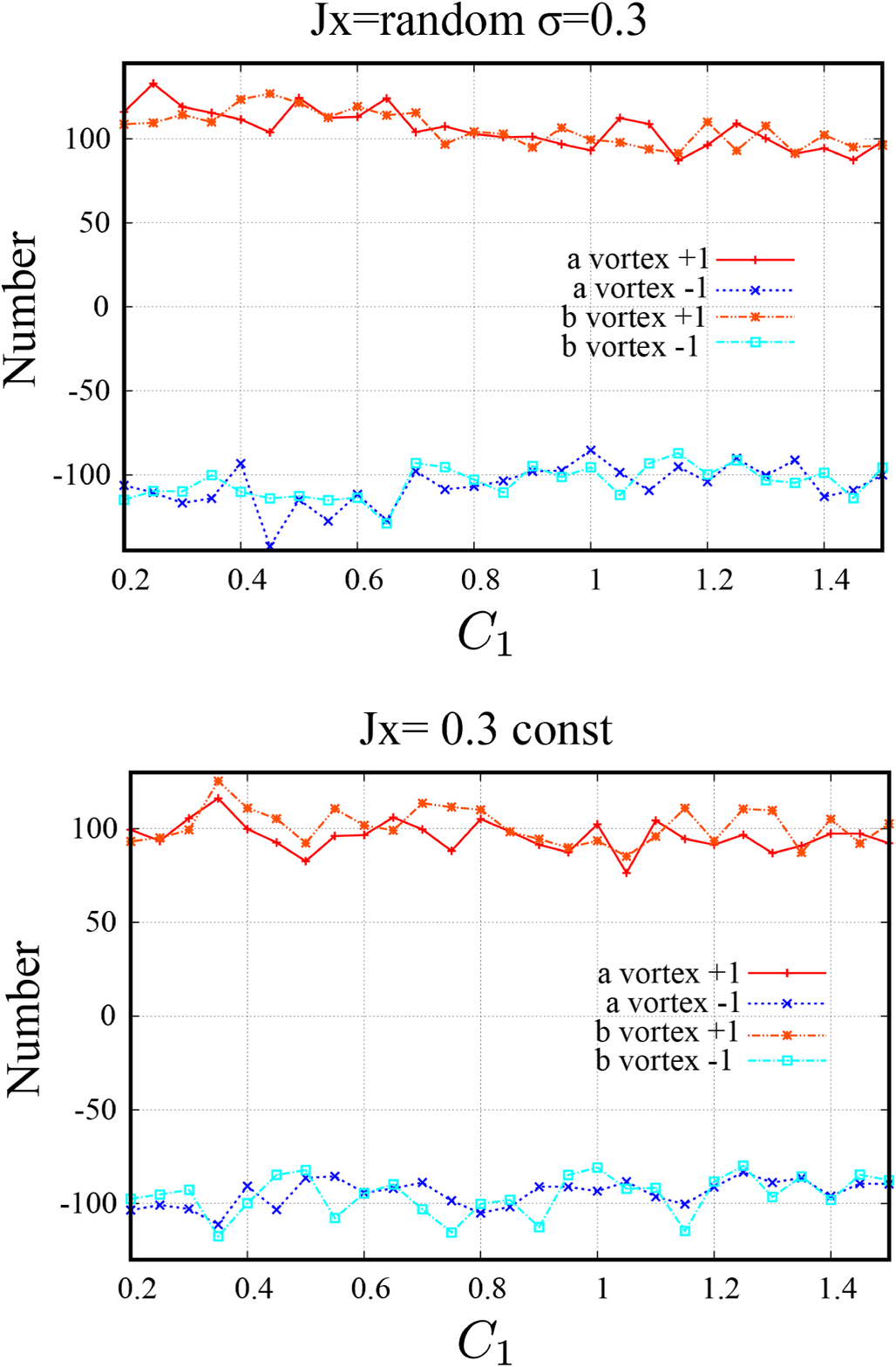} \hspace{0.5cm}
\includegraphics[width=6.5cm]{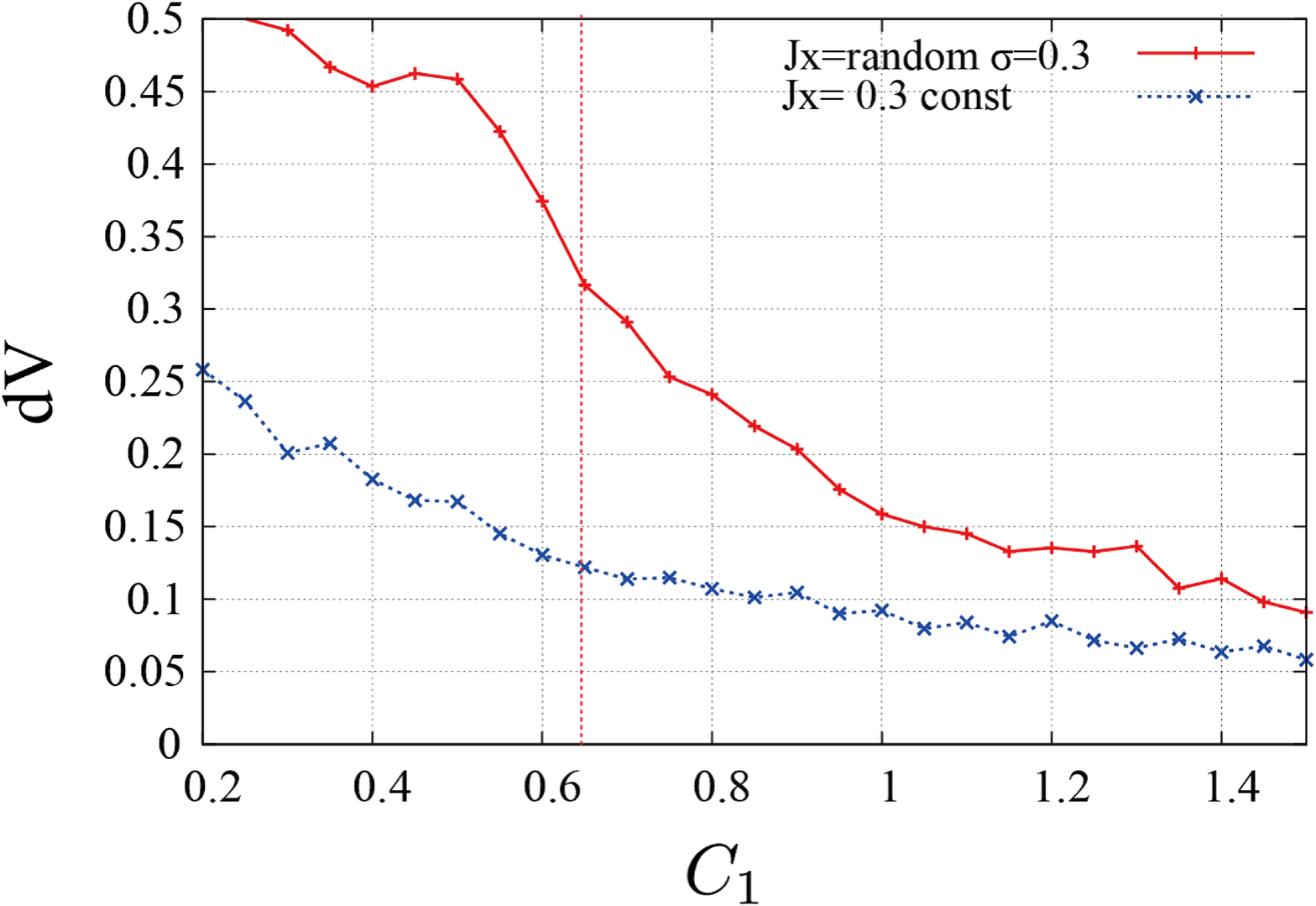}
\vspace{-0.3cm}
\caption{(Color online)
Calculations of vortex number (left) and $dV$ in Eq.(\ref{dV}) (right)
for $\sigma_x=0.3$ in the
random case and for constant $\tilde{J}_x=0.3$ with $\tilde{J}_y=0$.
Phase transition takes place at $C_1\sim 0.62$ for both the random
and constant $\tilde{J}_x$ cases.
Numbers of vortex and anti-vortex do not change through the phase
transition from the PM to FM /RFIO states, whereas $dV$ does.
Larger $dV$ indicates smaller energy of a vortex pair of the $a$- and $b$-atoms.
$C^a_3=C^b_3=0.3$ and $c_\tau=2.0$.
System size $L^2=16^2$.
}\vspace{-0.5cm}
\label{fig:dV}
\end{center}
\end{figure}
%%%%%%%%%%%%%%%%%%%%%%%%%%%%%%%%%%%%%%%%%%%%%%%%%%%%%%%%%%%

We calculate the local density of vortices $V^a_{r}$ and $V^b_{r}$,
which is defined as   
\begin{equation}
 V^A_{r}\equiv \left\{
               \begin{array}{cc}
              v_{Ar}, & |v_{Ar}|\geq1/2  \\
              0, &  |v_{Ar}|<1/2
             \end{array}
             \right.
\label{Vdensity}
\end{equation}
with the vorticity $v_{Ar}$ at site $r$ of the 3D space-time lattice,
\begin{eqnarray}
v_{Ar}&\equiv&{1 \over 4}\Big[\sin(\theta_{A,r+\hat{x}}-\theta_{A,r}) 
+\sin(\theta_{A,r+\hat{x}+\hat{y}}-\theta_{A,r+\hat{x}}) 
-\sin(\theta_{A,r+\hat{x}+\hat{y}}-\theta_{A,r+\hat{y}}) \nonumber  \\
&&-\sin(\theta_{A,r+\hat{y}}-\theta_{Ar})\Big],
\label{VAr}
\end{eqnarray}
where $A=a, b$.
Here we have introduce a cutoff and set
$V^A_{r}=0$ if $|v_{Ar}|$ is smaller than $1/2$.
This cutoff is useful for clarify locations of vortices.
From the local vortex density $V^A_r$ in Eq.(\ref{VAr}), we measure 
the overlap of the vortex configurations of the $a$- and $b$-atoms 
by calculating $dV$ in each time slice, which is defined as 
\begin{equation}
dV={1 \over N_v}\sum_{r\in \{\tau\mbox{:fixed}\}}
(V^a_r-V^b_r)^2,
\label{dV}
\end{equation}
where $N_v$ is the total number of vortex $N_v=N^{a+}_v-N^{a-}_v
\simeq N^{b+}_v-N^{b-}_v$,
$N^{A +}_v=\sum_{r\in \{\tau\mbox{:fixed}\}} 
V^A_r\theta(V^A_r)$, and
$N^{A -}_v=\sum_{r\in \{\tau\mbox{:fixed}\}} 
V^A_r\theta(-V^A_r)$ ($A=a,b$) with the Heaviside 
$\theta$-function, $\theta(x)$.
We show the calculation of $dV$ and also the total number of vortex and
anti-vortex, $(N^{A +}_v, \ N^{A -}_v)$ in Fig.\ref{fig:dV}.
Larger $dV$ means a smaller overlap of the $a$- and $b$-vortices.

Transition from the PM state to the RFIO state takes place
at $C_1\simeq 0.62$.
Density of vortex and anti-vortex does not change substantially in the PM
and RFIO phases, but the vortex overlap $dV$ changes drastically 
at the phase boundary.
See Fig.\ref{fig:dV}.
We also show similar quantities for the nonrandom case with constant $\tilde{J}^x_i=0.3$.
%and the genuine bosonic t-J model with $\tilde{J}^x_i=\tilde{J}^y_i=0$.
The result indicates that energy of the vortex pair in the RF system
is smaller than that in the system of constant $\tilde{J}^x_i=0.3$.
From the above result, it is also expected that
a shape of the brick wall between $a$- and $b$-vortex pair in the RFIO state is
different from that in the constant field, although it is not
so easily to observe it by snapshots of the MC simulations.
See Fig.\ref{fig:vortexp}.

%%%%%%%%%%%%%%%%%%%%%%%%%%%%%%%%%%%%%%%%%%%%%%%%%%%%%%%%%%%
%FIG.GP1
\begin{figure}[h]
\begin{center}
\includegraphics[width=8cm]{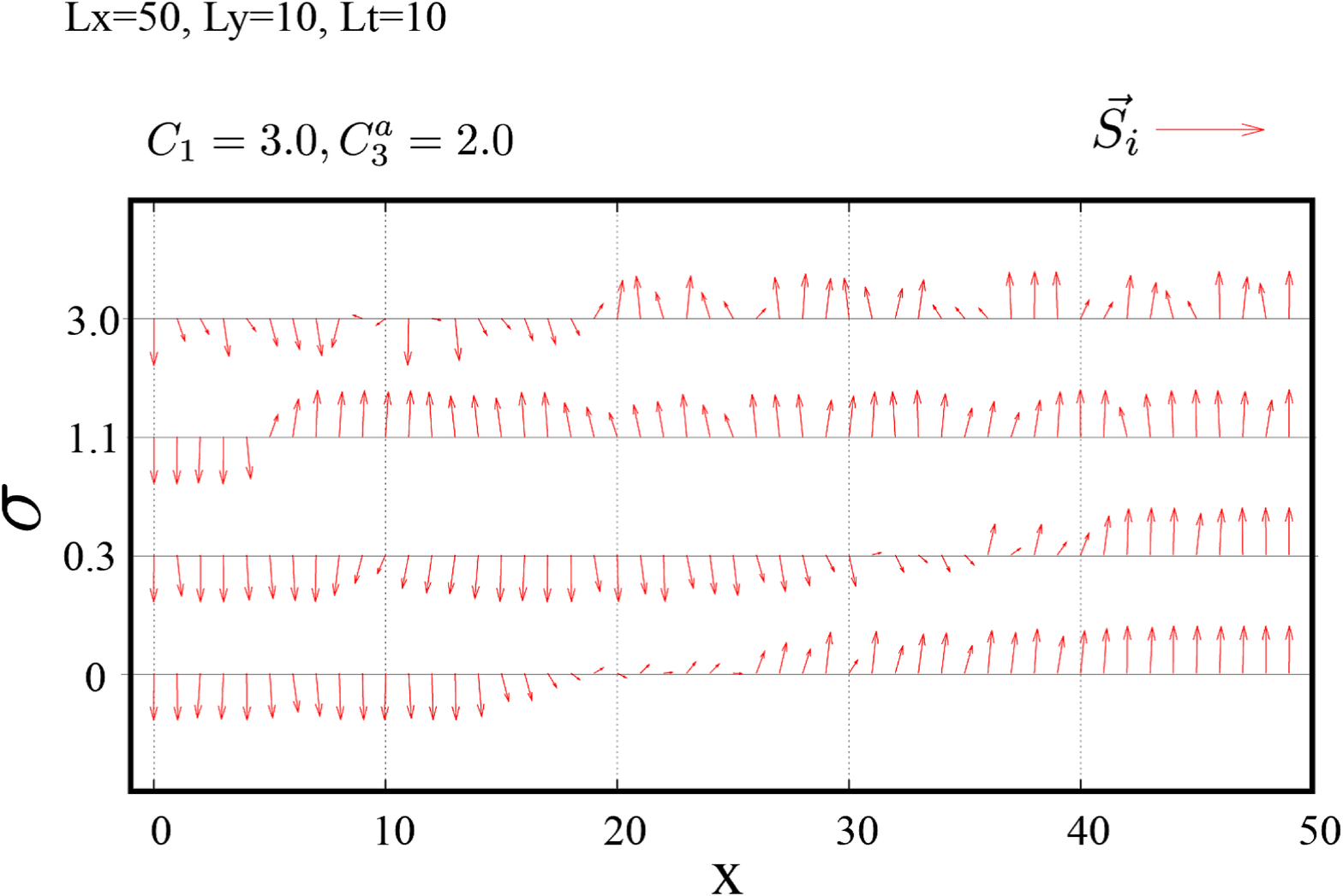}
\includegraphics[width=8cm]{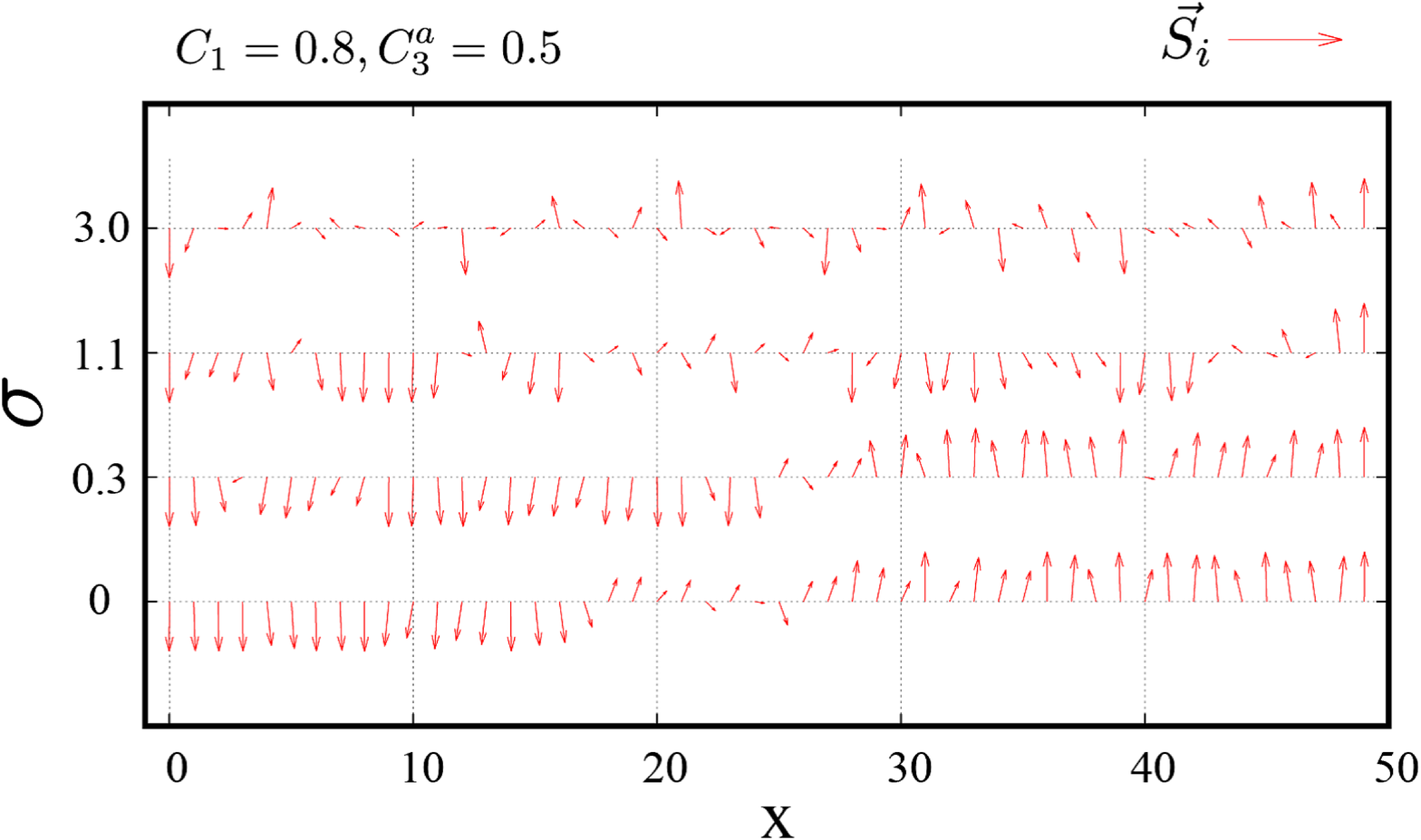}
\includegraphics[width=8cm]{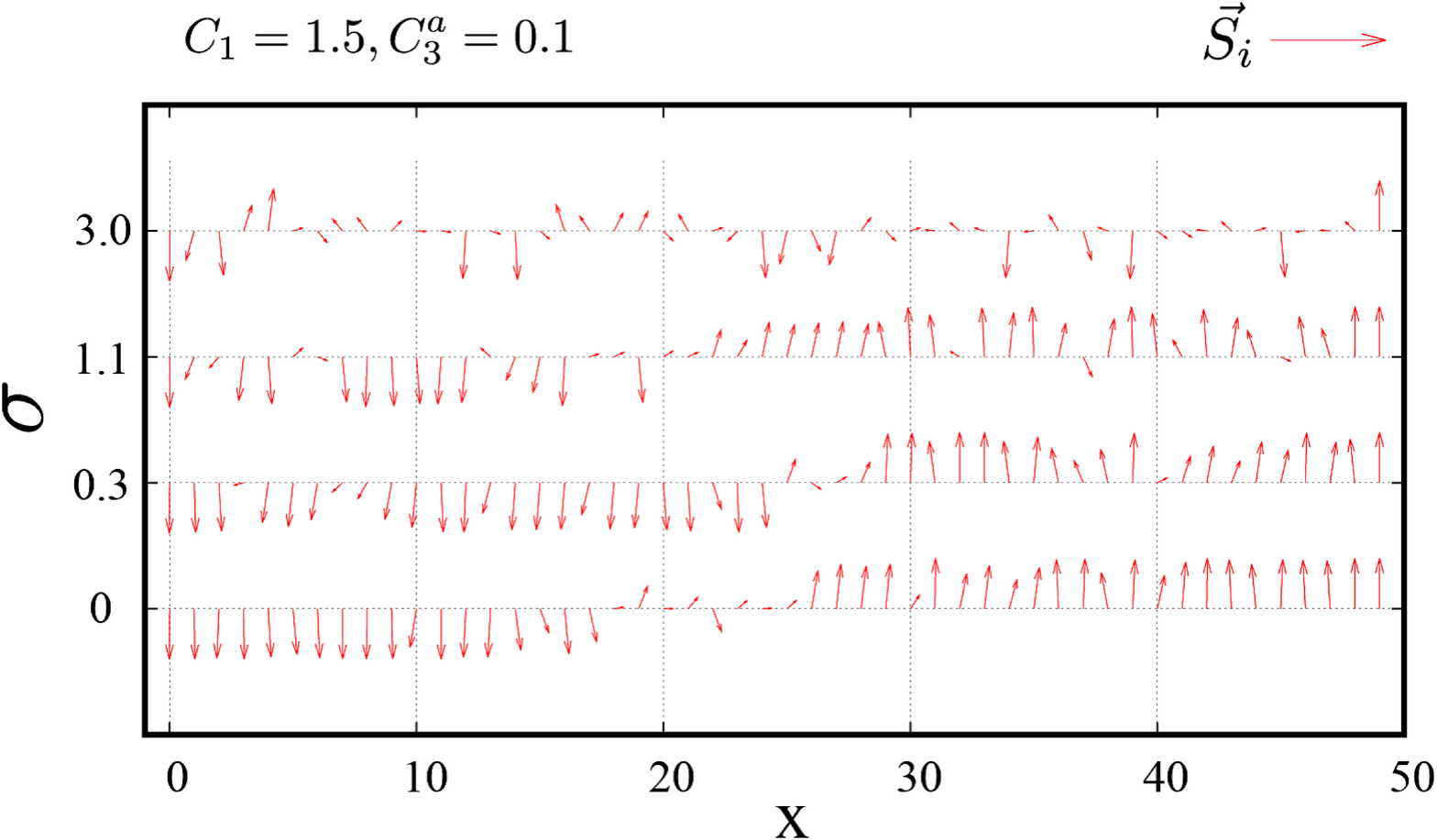}
\vspace{-0.3cm}
\caption{(Color online) (Top panel)
Expectation value of spin $\vec{S}_r$, 
$\langle \vec{S}_r \rangle$, for various $\sigma_x$ with the boundary condition 
such as $\uparrow (\downarrow)$ on the right (left) boundary.
Measured length of $\langle \vec{S}_r \rangle$ at each site is almost unity. 
In the region of kinks, length of $\langle \vec{S}_r \rangle$ is modified 
to show the direction of $\langle \vec{S}_r \rangle$ clearly.
For larger $\sigma_x$, the random variable $\tilde{J}_x$
fluctuates more strongly.
Result obviously shows that in the region of the moderate fluctuation of $\tilde{J}_x$,
$\sigma_x=0.3\sim 1.1$, the domain wall is rather thin compared to the case
of $\tilde{J}_x=0 \ (\sigma_x=0)$.
However, in the case of $\sigma_x=3.0$, the spins fluctuate rather strongly
as a result of a large fluctuation of $\tilde{J}_x$.
System size is $L_x=50, L_y=10$ and $L_t=10$.
$C_1=3.0$ and $C_3=2.0$.
(Middle and bottom panels) Similar behavior is observed near phase boundary
for $(C_1, C_3)=(0.8, 0.5)$ and $(C_1, C_3)=(1.5, 0.1)$.
}\vspace{-0.5cm}
\label{fig:DW}
\end{center}
\end{figure}
%%%%%%%%%%%%%%%%%%%%%%%%%%%%%%%%%%%%%%%%%%%%%%%%%%%%%%%%%%%

Let us turn to the domain wall of the relative phase $(\theta_{ar}-\theta_{br})$.
This domain wall is closely related to the ``string" connecting $a$-atom and $b$-atom
vortices.
See Fig.\ref{fig:vortexp}.
As we explained above, the calculation of $dV$ indicates that energy of the domain 
wall is getting smaller as the randomness of $\tilde{J}_x$ is getting larger from 
$\sigma_x=0$.
Then in the case of a large $\sigma_x$, the pseudo-spin loses its order as a result of 
a large spatial fluctuation of $\tilde{J}_x$.

To verify that the above expectation is correct, we investigate configurations 
generated by the boundary condition such that the spins $\vec{S}_r$ 
on the left spatial boundary have $(\theta_{ar}-\theta_{br})=-\pi/2$, 
whereas on the right spatial boundary $(\theta_{ar}-\theta_{br})=\pi/2$.
In Fig.\ref{fig:DW}, we show the expectation value of spin 
$\langle \vec{S}_r \rangle$ for various $\sigma_x$.
For the case $C_1=3.0$ and $C^a_3=C^b_3=2.0$,
the result obviously indicates that, from $\sigma_x=0$ to $\sigma_x=1.1$, 
the stiffness of spins is getting stronger
as a result of larger fluctuation of the random variable $\tilde{J}_x$.
On the other hand in the case of $\sigma_x=3.0$, the pseudo-spins 
fluctuate rather strongly.
Similar behavior is observed in the other cases in Fig.\ref{fig:DW}.
This indicates that the strongly fluctuating random-external field 
destroys the spin order.
Then,
it is an interesting problem to determine a critical randomness $\sigma_c$
for the order-disorder phase transition observed in the present numerical
simulations.

%%%%%%%%%%%%%%%%%%%%%%%%%%%%%%%%%%%%%%%%%%%%%%%%%%%%%%%%%%%%%
\section{Conclusion}

In this paper, we studied effect of a ``random external field" on the phase
diagram of the bosonic t-J model, properties of the states and the
low-energy excitations in the RFIO state.
This external field is realized by a random Rabi oscillation between two 
internal states in an atom induced by a random Raman laser.
In the phase diagram of the bosonic t-J model without the random field, 
there exist ordered states such as the pseudo-spin FM state and the 2SF.
We first investigated how the phase diagram is changed as a result of the
random field and found that the ordered states move to the states with the RFIO.
In the RFIO states, the original U(1) symmetry reduces the $Z_2$-Ising type,
and therefore low-energy excitations in the RFIO states have different
propertied from those in the original ordered states of the t-J model.

By the replica-MFT, we first studied the low-energy properties of the quantum spin 
system in a random external field, and found that, for a sufficiently strong randomness,
there appear the preferred directions of the spin order, which are perpendicular to
the applied field. 
Then, by using the MC simulations, we studied the phase diagram of the effective
field theory of the t-J model in applied random external fields, 
$(\tilde{J^x}, \tilde{J}^y)$.
We found that the direction of the spin order is determined by which
component of the applied field, $(\tilde{J^x}, \tilde{J}^y)$, is larger.
We also studied the finite-$T$ phase diagram and found that the RFIO of the spin 
survives at intermediate temperatures although the SF is destroyed by 
the thermal fluctuations. 

Finally, physical properties of topological excitations such as the vortex  
and domain wall were studied.
Binding energy of the vortex pair of the $a$- and $b$- bosons is smaller in the
RFIO compared to that in the genuine t-J model.
This means that average distance between $a$- and $b$-vortices in a single vortex pair
is getting longer as $\sigma_x$ increases.
Similarly, the width of the domain wall is thiner in the RFIO state.
We hope that the above findings are observed by experiments on cold atomic
gases.
In near future, we shall report studies on the behavior of vortex lattices that form
as a result of the coupling to an artificial external vector potential.

%%%%%%%%%%%%%%%%%%%%%%%%%%%%%%%%%%%%%%%%%%%%%%%%%%%%%%%%%%%%%

\end{document}